\documentclass[useAMS,usenatbib]{mn2e}
\usepackage[pdftex]{graphicx}
\usepackage{aas_macros}
\usepackage{setspace}
\usepackage{url}
\usepackage{deluxetable}
\usepackage{setspace}
\usepackage{natbib}
\usepackage{multicol}
\usepackage{aas_macros}
\usepackage{amsmath}
\usepackage{amssymb}
\usepackage{pdflscape}
\usepackage{bm}

\newcommand {\apgt} {\ {\raise-.5ex\hbox{$\buildrel>\over\sim$}}\ }

\title[High Velocity-dispersion Cold Gas in ULIRG Outflows. I.]{High Velocity-dispersion Cold Gas in ULIRG Outflows. I: Direct Simulations}
\author[D. J. Williamson, Robert J. Thacker, E. Scannapieco, and M. Br\"{u}ggen]{D. J. Williamson$^{1,2}$\thanks{E-mail:
david-john.williamson.1@ulaval.ca}, Robert J. Thacker$^{3}$, E. Scannapieco$^{4}$, and M. Br\"{u}ggen$^{5}$\\
$^{1}$D\'{e}partement de physique, de g\'{e}nie physique et d'optique, Universit\'{e} Laval, Qu\'{e}be c, QC, G1V 0A6, Canada\\
$^{2}$Centre de Recherche en Astrophysique du Qu\'{e}bec, Qu\'{e}bec, QC, Canada\\
$^{3}$Department of Astronomy and Physics, St Mary's University, Halifax, NS, B3H 3C3, Canada\\
$^{4}$School of Earth and Space Exploration, Arizona State University, Tempe, AZ, 85287-1404, USA\\
$^{5}$Hamburger Sternwarte, Gojenbergsweg 112, D-21029 Hamburg, Germany\\
}
\begin{document}

\date{This pre-print submitted Mar 2014 (Accepted to MNRAS, Mar 2014) }

\pagerange{\pageref{firstpage}--\pageref{lastpage}} \pubyear{2013}

\maketitle

\label{firstpage}

\begin{abstract}
Observations have revealed cold gas with large velocity dispersions ($\approx300$ km/s) within the hot outflows of ultra-luminous infrared galaxies (ULIRGs). This gas may trace its origin to the Rayleigh-Taylor (RT) fragmentation of a super-bubble or may arise on smaller scales. We model a ULIRG outflow at two scales to recreate this gas in three-dimensional hydrodynamic simulations using FLASH. Although resolution is limited, these models successfully produce cold gas in outflows with large velocity dispersions. Our small-scale models produce this cold gas through RT fragmentation of the super-bubble wall, but the large-scale models produce the cold gas after hot bubbles fragment the disc's gas into cold clouds which are then accelerated by thermal pressure, and supplemented by cooling within the outflow. We produce simple mock spectra to compare these simulations to observed absorption spectra and find line-widths of $\approx 250$ km/s, agreeing with the lower end of observations.
\end{abstract}


\section{Introduction}
Nuclear and galactic scale outflows are common in ultra-luminous infrared galaxies (ULIRGs), and have been detected in X-Ray  \citep{1996ApJ...457..616H,2003ApJ...591..154M,2003ApJ...592..782P,2009ApJ...691..261T} and H$\alpha$ \citep{2004ApJ...602..181C} emission as hot lobes extending $10-15$ kpc beyond the infrared-luminous portion of the galaxy. These ``super-winds'' are primarily driven by supernovae, with outflow rates comparable to the host's star formation rate ($\approx10-1000 M_\odot$/yr), and correspondingly high luminosities ($10^{41}-10^{44}$ erg/s) and projected velocities ($300-400$ km/s).

Several observations \citep[e.g.][]{1993AJ....105..486P,2000ApJS..129..493H,2005ApJ...621..227M,2006ApJ...647..222M} have detected cold gas with large velocity dispersions in these outflows. The presence of this gas provides a challenge to theoretical models, which must explain how such a cold component \citep[i.e. NaI and KI absorption lines, with ionization potentials of 5.1 eV and 4.3 eV,][]{2005ApJ...621..227M} can exist within a flow of very hot ($T\sim10^6$ K) gas. These models must also explain the large velocity dispersion of this gas --- i.e. the large non-thermal broadening of the NaI lines. A number of works have investigated explanations for the cold component, often supported by numerical calculations \citep{1986PASJ...38..697T,2005ApJ...618..569M,2009ApJ...698..693F,2010ApJ...713..592E} and simulated spectra \citep{2009ApJ...698..693F,2011ApJ...734...24P}. In these models, cold gas is either produced in the disc and then advected by ram-pressure, or produced by gas rapidly cooling within the wind through radiative processes.

In the simulations of \citet{2009ApJ...698..693F}, cold gas is produced by turbulence in the wind, which leads to dense condensations that rapidly cool. Specifically, these condensations trace their origin to a super-bubble, which is inflated by the outflow, driving a ``snowplough'' that builds up a dense bubble wall. The density of this wall allows it to cool efficiently, due to the $\rho^2$ dependency in radiative cooling, and the wall is supported against gravity by the pressure of the hot low-density gas within it. This situation is extremely susceptible to Rayleigh-Taylor (RT) \citep{1950RSPSA.201..192T,1984PhyD...12....3S} and Richtmyer-Meshkov (RM) \citep{CPA:CPA3160130207,meshkovref} instabilities, which cause the bubble wall to fragment into clouds that become a cold high velocity-dispersion component of the wind.

This situation is difficult to analyze numerically, because the RT and RM instabilities can be strong at a wide range of wavelengths, and so simulations that strongly depend on the RT and RM instabilities will not converge until a fine enough resolution is used to resolve the full turbulent cascade. Alternatively, a sub-grid model for turbulent evolution can be considered, as we do in part II of this series.

To better resolve the scales relevant to turbulent formation and destruction of clouds, the simulations of \citet{2009ApJ...698..693F} were performed in two dimensions with cylindrical symmetry. While this allows a resolution of as fine as $0.1$ pc in a $100$ pc by $200$ pc box, it suppresses modes of instability and gas flow that may be present in three dimensions. Furthermore, even at this high resolution, \citet{2009ApJ...698..693F} note a significant resolution dependence in the scale of these clouds, which suggests that the turbulence is still not fully-resolved.

There is thus motivation to reexamine this scenario with a fully three-dimensional hydrodynamic model. However, even using adaptive methods, we are not able to attain an equivalent resolution in 3D models on currently available computer resources. Instead we develop 3D models with two scales of size --- one with the same scale as \citet{2009ApJ...698..693F} but lower resolution, and a full-scale galaxy model to examine large-scale effects. We also vary the initial conditions and mass loading rates, and thus produce a suite of models to investigate numerical effects. To facilitate a more direct comparison with observations, we have also produced a raytracing code for calculating mock NaI absorption spectra of the models.

The layout of the paper is as follows: in section~\ref{simmodel} we give a summary of the simulation method for both the full galaxy and galaxy centre models. In section~\ref{raytracespectra} we present the raytracing and mock spectra algorithms. In section~\ref{cloudfind} we summarise the cloud tracking technique. In section~\ref{resultscoldflows} we present our results for the full galaxy models (section~\ref{fullgalaxyresults}) and galaxy centre models (section~\ref{galaxycentreresults}). We present our conclusions in section~\ref{coldflowsconc}.

\section{Simulation model}\label{simmodel}

\subsection{Initial Conditions}\label{ic_section}

The initial conditions were generated using the galactic disc generator of \citet{2006MNRAS.369..567R,2010MNRAS.405.1634S} with some modifications. All of our initial conditions were produced by this generator, although the full galaxy and galaxy centre models are generated with different choices of parameters.

In this model, only the gaseous component is explicitly evolved, while the dark matter component is represented by a fixed gravitational potential, and the stellar component is modelled by both a gravitational potential and a feedback algorithm. The gaseous disc is initially axisymmetric, following a $\mathrm{sech}$ profile in both the radial and vertical directions:

\begin{equation}
\rho(r,z)=\frac{M_\mathrm{gas}}{4 C \pi^2 h_r^2 h_z}  \mathrm{sech}(r/h_r)\,\mathrm{sech}(|z|/h_z)+\rho_\mathrm{amb},
\end{equation}
where $M_{gas}$ is the total gass mass, $h_r$ and $h_z$ are the radial and vertical scale heights, $\rho_\mathrm{amb}$ is the ambient density, and $C$ is Catalan's constant, defined by $2C = \int_0^{\infty} {x\,\mathrm{sech}(x) \mathrm{d} x}$.

Pressure is set to provide vertical hydrostatic equilibrium by

\begin{equation}
P(r,z)=P_\mathrm{amb}-\int_{z_\mathrm{max}}^z \rho(r,z) g(r,z) \mathrm{d}z,
\end{equation}
where $P_\mathrm{amb}$ is the ambient pressure, $z_\mathrm{max}$ is the height of the computational volume, and $g(r,z)$ is the acceleration due to gravity, given by a combination of a halo and a thick-disc potential.

As in \citet{2010MNRAS.405.1634S} we use the halo potential defined in \citet{1995ApJ...447L..25B} and given explicitly in \citet{2000ApJ...538..559M}:

\begin{equation}
\begin{split}
\phi_\mathrm{halo} = -\pi G \rho_{d0} r_{d0}^2 & \Big\{ -2\left(1+\frac{r_{d0}}{R}\right)\mathrm{atan}\left(\frac{r_{d0}}{R}\right)\\
&+2\left(1+\frac{r_{d0}}{R}\right)\ln\left(1+\frac{R}{r_{d0}}\right)\\
&-\left(1-\frac{r_{d0}}{R}\right)\ln\left[1+\left(\frac{R}{r_{d0}}\right)\right] \Big\},
\end{split}
\end{equation}
where $r_{d0}$, and $\rho_{d0}$ are the parameters that define the spatial and mass scale of the halo, and $R$ is the radius in spherical coordinates. This potential is a phenomenological model based on observations of dwarf galaxies, where the dark matter density profile can be more directly probed. It is designed to resemble an isothermal profile at low radii, preventing the problem of ``cuspiness'', while agreeing with the commonly used Navarro-Frenk-White profile at large radii \citep{1996ApJ...462..563N}. 

The disc potential we use is a Plummer-Kuzmin disc \citep{2008gady.book.....B}:
\begin{equation}
\phi_\mathrm{disc} = \frac{-G M_\mathrm{disc}}{\sqrt{r^2+[h_r+(z^2+h_z^2)^{1/2}]^2}},
\end{equation}
where $M_\mathrm{disc}$ is the mass of the disc, $h_r$ and $h_z$ are the radial and vertical scale heights, and $r$ and $z$ are position in cylindrical coordinates. Given that our initial conditions do not account for the complex irregular structure of a ULIRG, we do not include a bulge potential as this would not significantly improve our model.

To break the symmetry of the initial conditions, we have also produced models that included asymmetric density perturbations. This is only applied to some of the galaxy centre simulations, as the full galaxy models naturally evolve asymmetrically through a stochastic feedback algorithm (section.~\ref{feedbacksection}). Using the cartesian coordinate system $(x,y,z)$ where $(x,y,z)=(0,0,0)$ is the centre of the feedback region and $(x,y,0)$ defines the plane of the ULIRG disc, the perturbation is applied by defining factors $\xi$ and $f$ as

\begin{equation}
\begin{split}
\xi = & [1+A_\mathrm{lumpy}\cos(\pi(x+\delta)/\lambda)\cos(\pi(y+\delta)/\lambda)]^2\\
	  \times&[1-A_\mathrm{lumpy}\cos(\pi(x+\delta)/\lambda)\cos(\pi(y+\delta)/\lambda)]^2\\
\end{split}
\end{equation}
and

\begin{equation}
f =  [1-(1-\xi) \mathrm{erfc}(-y/10h_z)/2],
\end{equation}and defining the perturbed density $\rho^\prime(x,y,z)$ by
\begin{equation}
\rho^\prime (x,y,z) = \rho(x,y,z) f^{\cos[\pi(z+\delta)/\lambda]}.
\end{equation}Here $\lambda$ is the wavelength of the perturbation, $\delta$ is the offset to break symmetry, the complementary error function (erfc) is a term that smoothly reduces the amplitude of the perturbation towards zero over several scale heights, $A_\mathrm{lumpy}$ is the amplitude of the perturbation, and the cosine term in the $z$-direction is placed in the exponential in order for the perturbation to be significant when compared to the rapid exponential rate of change of density in the vertical direction. This provides an approximately sinusoidal perturbation that is dampened as $z$ increases, which prevents extreme temperature variations in the sparse higher-altitude gas. To further reduce symmetry, this perturbation is applied to the density twice, using $\lambda=37.3$ pc and $\delta=0$ pc in the first instance, and $\lambda=97.2$ pc and $\delta=22.7$ pc in the second, with $A=0.5$ in both cases. These values are not intended to closely match the details of the density field in the central region of a real ULIRG, but are chosen such that the wavelength of the perturbation is large enough to be well-resolved, but still small enough that at least one wavelength can fit within the simulation domain. The intention here is only to ensure that any structure that is produced in the simulation does not directly result from the symmetry of the initial conditions and feedback algorithm.

\subsection{Galaxy centre model}
The simulation domain of the galaxy centre model is a cartesian grid of size $200^3$ pc$^3$. We reduce the computational load by using reflective boundary conditions across the $z=0$ plane -- which is justified due to the symmetry of the flow in this situation. Following \citet{2009ApJ...698..693F}, the disc has a vertical scale-height of $7$ pc, a radial scale-length of $700$ pc, and a mass of $10^{10} M_{\odot}$, which represents only the central molecular disc.

As the scale-length is much larger than the box size, this produces a disc that has almost no radial density variation within the box. The disc is also truncated using the complementary error function to ensure the density at the edge of the box is small, and large amounts of gas do not rotate out of the box. As this truncation is much shorter than the radial scale-length, it does not accurately represent the density profile, and the simulation is stopped when the inflated bubble reaches the truncated region.

\citet{2009ApJ...698..693F} used a range of luminosities from $10^{41}$--$10^{43}$ erg/s. In our model we set $\dot{M}_*=50\mathrm{M_\odot /yr}$, which is equivalent to a luminosity of $6.3\times 10^{42}$ erg/s, on the upper edge of their energy range. This results in a mass-loading rate of $2$ $\mathrm{M_\odot /yr}$. The parameters for the galaxy centre models are summarized in the top section of Table~\ref{flashmodels}.
\begin{center}
\begin{table}
\begin{tabular}{ c c c c }
\hline\hline
Name & $M_\mathrm{SN}$ (M$_\odot$) & $l_\mathrm{ref}$ & $A_\mathrm{lumpy}$ \\
\hline
SVLowRes & 6 & 1 &  0.\\
SLowResMassless & 0 & 2 &  0.\\
SLowResLight & 4 & 2 &  0.\\
SLowRes & 6 & 2 &  0.\\
SLowResHeavy & 16 & 2 &  0.\\
SMedRes & 6 & 3 &  0.\\
SHighRes & 6 & 4 &  0.\\
SLowResLumpy & 6 & 2 &  0.5\\
SMedResLumpy & 6 & 3 &  0.5\\
SHighResLumpy & 6 & 4 &  0.5\\
\hline\hline
BMedRes & 6 & 2 &  0.\\
BHighRes & 6 & 3 &  0.\\
\hline
\end{tabular}
\caption{\label{flashmodels} Parameters in disc outflow models. The prefix ``S'' refers to the ``small'' central-disc models, while the prefix ``B'' refers to the ``big'' full-scale galaxy models. $M_\mathrm{SN}$ is the mass ejected in each supernova (i.e. the mass-loading), $l_\mathrm{ref}$ is the number of refinement levels (including the top level), and $A_\mathrm{lumpy}$ gives the amplitude of perturbations to the initial density distribution.} 
\end{table}
\end{center}

\subsection{Full galaxy model}

The full galaxy model is simulated within a cartesian grid of size $240^3~\mathrm{kpc}^3$. The vertical asymmetry of the feedback does not permit a $z=0$ reflective boundary condition for this model.

This disc is intended to be an isolated simple ULIRG prototype. As ULIRGs are obscured by dust and locally rare, their characteristics are poorly constrained by observations. However, we still need to choose parameters that are consistent with observations and previous models. Direct observations have provided estimates for the star formation rates (SFRs), molecular gas masses, and dynamical masses of these objects \citep{2008MNRAS.391..420S,2008ApJ...677..957F,2011ApJ...737...64A}. Sub-millimetre galaxies (SMGs) and dust-obscured galaxies (DOGs) may also be high-redshift analogues for ULIRGs, and observations of these objects provide additional constraints \citep{2005MNRAS.356.1191B,2009MNRAS.400.1919N,2010MNRAS.407.1701N}. These observational data can be supplemented by the parameters used in previous simulations of ULIRG progenitor mergers \citep{2006MNRAS.371..805B,2009PASJ...61..481S,2010MNRAS.404.1355D,2012ApJ...746...26M}.

Informed by these observations and simulations, we employ a gaseous disc with mass of $1.0\times10^{10} M_\odot$, scale-length of $8$ kpc, and scale-height of $2$ kpc. The gas disc is superimposed on a background potential equivalent to a stellar disc of mass $4.0\times10^{10} M_\odot$ with a scale-length of $16$ kpc and a scale-height of $4$ kpc, as well as a dark-matter halo potential of mass $3.0\times 10^{12} M_\odot$ and scale-radius $100$ kpc.

We performed test simulations with a SFR chosen according to Kennicutt-Schmidt law \citep{1998ApJ...498..541K}, but these simulations did not produce outflows due to the well-known problem of over-cooling at limited resolution. Solutions to this problem include using an effective density in the cooling of heated particles \citep{2000ApJ...545..728T}, shutting off cooling temporarily in heated cells \citep{1997A&A...325..972G}, directly applying momentum ``kicks'' in addition to thermal energy \citep{1993MNRAS.265..271N,1994ApJ...437..611M,2001ApJ...557..605S,2003MNRAS.339..289S}, and making use of a sub-grid turbulence model \citep{2010MNRAS.405.1634S}. In part II of this paper series we will include a sub-grid model, but for this work we compensate for over-cooling by simply setting an SFR above the Kennicutt-Schmidt law. We also note that it has been argued that ULIRG SFRs lie above the global Kennicutt-Schmidt law \citep{2000ApJ...537..613A,2010ApJ...714L.118D,2011A&A...533A..19B}. Hence we set the star formation rate to $200 M_\odot$/yr, which while at the lower end of ULIRG SFRs, is greater than the Kennicutt-Schmidt SFR for our galaxy model's surface density. The parameters for the galaxy-scale models are summarized in the bottom section of Table.~\ref{flashmodels}.

\subsection{Cooling Function}

For this code, we make use of the cooling curves, $\Lambda(T)$, of \citet{1976ApJ...204..290R} and \citet{1986RvMP...58....1S}. We set the metallicity to $Z=0.05Z_\odot$. The cooling algorithm changes the internal energy $e_i$ of each cell by applying $\Delta e_i = \Lambda n^2 \Delta t_j$, several times each time-step, with the cooling time-step $\Delta t_j$ set such that $|\Delta e_i|<|e_i|/10$ in each step, and that $\sum \Delta t_j$ is equal to the true time-step.

\subsection{Feedback from Star Formation}\label{feedbacksection}

We use the feedback model implemented by \citet{2010MNRAS.405.1634S}, with some modifications. In this model, the number of supernovae is tallied over time. Informed by the cosmic SN rate \citep{2004ApJ...613..189D,2004ApJ...600L.103G} and by initial mass functions, it is assumed that $150 $M$_\odot$ of star formation is required to produce a single supernova \citep[e.g][]{2002ApJ...574..590S}. When the cumulative count of supernovae exceeds a certain number, a bubble of hot gas is produced in the disc, and these supernovae are decremented from the total supernova count. The number of supernovae required to produce a bubble ($N_\mathrm{SN}$) is determined by generating a random variable $\zeta\in[0,1]$ and calculating

\begin{equation}
N_\mathrm{SN} = \frac{N_\mathrm{SN,min}}{1-\zeta},
\end{equation}
where $N_\mathrm{SN,min}$ is a model parameter. If $N_\mathrm{SN}>N_\mathrm{SN,max}$, where $N_\mathrm{SN,max}$ is another model parameter, then $\zeta$ is regenerated until $N_\mathrm{SN}\le N_\mathrm{SN,max}$. $N_\mathrm{SN,min}$ and $N_\mathrm{SN,max}$ determine whether feedback consists of a small number of large bubbles, or a large number of small bubbles. The radius of the bubble is set such that it is at least the size of the region containing twice the mass in gas of the mass converted into stars, as well as being greater than a resolution-dependent minimum size that ensures that the bubble covers at least one cell.

The feedback is implemented by calculating the mass density and specific internal energy associated with the bubble,
\begin{equation}
\rho_\mathrm{SN} =  M_\mathrm{SN} \frac{N_\mathrm{SN}}{4 \pi R_\mathrm{bub}^3},\\
\end{equation}and
\begin{equation}
e_i,\mathrm{SN} =  \frac{N_\mathrm{SN}}{4 \pi R_\mathrm{bub}^3} f_\mathrm{SN} 10^{51} \mathrm{erg},
\end{equation}
and adding this energy and density to all gas cells with centres within $2 R_\mathrm{bub}$ by applying the transforms

\begin{equation}
\rho \mapsto \rho + \rho_\mathrm{SN}~\mathrm{erfc}\left(\frac{\vec{r}_\mathrm{cell}-\vec{r}_\mathrm{bub}}{0.05 R_\mathrm{bub}}\right),
\end{equation}and
\begin{equation}
e_{i} \mapsto \left[e_{i} \rho_\mathrm{old}+e_\mathrm{SN}\rho_\mathrm{SN}~\mathrm{erfc}\left(\frac{\vec{r}_\mathrm{cell}-\vec{r}_\mathrm{bub}}{0.05 R_\mathrm{bub}}\right) \right]/\rho_\mathrm{new}.
\end{equation}
Here, the erfc term smoothes the edges of the bubble and ensures that the only sharp discontinuities in density and temperature are those caused by the evolution and interaction of bubbles, and not those directly imposed by the feedback algorithm. Again this is a modification to \citet{2010MNRAS.405.1634S}'s algorithm.

In our fiducial models we set $f_\mathrm{SN}=0.6$, and $M_\mathrm{SN}=6 \mathrm{M_\odot}$, which allows us to parameterize the feedback's luminosity, $L_\mathrm{FB}$, in addition to the its mass loading rate, $\dot{M}$, in terms of the star formation rate $\dot{M}_*$, through the following equations:

\begin{equation}
L_\mathrm{FB} \approx 1.3\times 10^{43} \mathrm{erg s}^{-1} \left(\frac{\dot{M}_*}{100 \mathrm{M}_\odot \mathrm{/yr}}\right),\\
\end{equation}and
\begin{equation}
\dot{M} =\left(\frac{\dot{M}_*}{150 \mathrm{M}_\odot/M_\mathrm{SN}}\right) = \left(\frac{\dot{M}_*}{18.75}\right).
\end{equation}
Mass loading rates were varied in some models by altering $M_\mathrm{SN}$. Some unrealistically large values are included to investigate the effects of very large mass-loading rates.

\subsubsection{Feedback in full galaxy models}

In the full galaxy models, each bubble is placed randomly, with positions weighted according to $\rho_0(r,z)^{3/2}$ to reproduce the Schmidt Law, where $\rho_0(r,z)$ is the initial analytic distribution of density. Specifically, three random numbers $\xi_1,\xi_2,\xi_3 \in [0,1]$ determine the position of the bubble. Defining again a cartesian coordinate system $\vec{r}=(x,y,z)$ where the disc lies in the $(x,y)$ plane, the position is given by

\begin{align}
x =&~r \cos (2\pi \xi_1), \\
y =&~r \sin (2\pi \xi_1),
\end{align}
and
\begin{equation}
z = h_z \ln (\xi_3),
\end{equation}
where $r$ is determined by iteratively solving the equation
\begin{equation}
r=\log[(1+r)/(1-\xi_2)].
\end{equation}We have modified this model from that in \citep{2010MNRAS.405.1634S} to allow multiple bubbles to be produced in a single time-step, which is necessary due to the large star formation rate.

\subsubsection{Feedback in central disc models}

In the central disc models, the outflow is modelled as a single starburst. While $N_\mathrm{SN}$ is calculated as above, the bubbles are always placed in the centre of the disc at $z=0$ pc, and have a constant radius of $R_\mathrm{bub}=10$ pc. This approach is similar to that of \citet{2009ApJ...698..693F}, and allows a more direct comparison of the important difference in the models: that our simulations are performed in three dimensions instead of two (at the cost of lower resolution).

\section{Raytracing and Simulated Spectra}\label{raytracespectra}

To compare our results with observations, we developed a code to perform ray-tracing on FLASH checkpoint data, and obtain a synthetic absorption spectrum from the output. This spectrum covers the NaI line, as this is a useful tracer of cold gas. In particular, \cite{2005ApJ...621..227M} produced spectra for a sample of $z\sim1$ ULIRGs, and we use these as a basis for comparison.

\subsection{Raytracing Algorithm}

Our raytracing algorithm tracks a ray or a grid of rays with arbitrary angles through data from a FLASH checkpoint file. Various quantities can be integrated along a ray as it propagates through the FLASH adaptive mesh. By using a grid of rays, a column density or optical depth plot can be produced from arbitrary angles, or (as described below) the effects of the finite spatial resolution of a spectrometer can be emulated. Our approach follows a ray through the FLASH oct-tree refinement structure without converting the data to a uniform grid.

\subsection{NaI Line spectrum}\label{sodiumline}

\subsubsection{General process}
Our method for modelling the NaI line spectrum incorporates a number of assumptions. Firstly, we have assumed uniform background illumination by weighting the grid of rays evenly when performing an average. Secondly, we have not accounted for dust, which should produce a significant attenuation in ULIRGs, although when examining absorption lines this may be less important. Finally, we have ignored any spectral broadening from the Earth's atmosphere or from the instrument performing the observations; but given the limited physical resolution of the simulations, it is not necessary to model these effects in detail.

To model the NaI line, we assume a constant HI/NaI ratio, $N_\mathrm{NaI}=1.22\times10^6$, taken from \cite{2005ApJ...621..227M}. Following \citet{2009ApJ...698..693F}, we assume that any cell with a temperature below $5\times10^4$ K consists of cold, NaI-absorbing gas. As justified below, we also make a density cut at $10^{-21} \mathrm{g/cm}^3$, with the assumption that any gas above this density will contain some regions that have cooled sufficiently to absorb the NaI doublet. We assume that any gas which fulfils either of these criteria consists of NaI absorbing gas.

The line is calculated by integrating the optical depth through a ray across a range of wavelengths. The optical depth is summed for each cell the ray passes through, ignoring any emissivity, and the continuum intensity is normalised to $1$, so the final intensity of each frequency is given by
\begin{equation}
I(\nu)=\exp(-\tau_\nu)=\exp\left(-\sum \tau_{\nu,i} \right),
\end{equation}
where $\tau_{\nu,i}$ is the optical depth at frequency $\nu$ through cell $i$. This is calculated from \cite{1978ppim.book.....S}
\begin{equation}
\tau_{\nu,i} = N_i \alpha \lambda \Phi(\nu),
\end{equation} where the frequency-integrated cross-section for the interaction ($\alpha$) is given by $\alpha=0.015924$ cm$^2$ s, $N_i$ is the NaI column density, and the doppler profile $\Phi(\nu)$ is given by
\begin{equation}
\Phi(\nu)=\frac{1}{\sqrt{\pi b}} \exp(-v^2/b^2),
\end{equation}
where $b=\sqrt{2kT_i}$, $k$ being the Boltzmann constant and $T_i$ being the cell's temperature.
The optical depth is binned in $1000$ bins, representing line of sight velocities of $-500$ through $500$ km/s. Each optical depth bin is summed through a ray's path, and converted to intensity in the final step. Results for this method are given in sections \ref{specnoturb}, and \ref{specsmall}

\subsubsection{Spectral and Spatial Resolution}

The ray-tracing procedure produces a line profile for an infinitely narrow beam passing through the system. This represents the limit where an instrument can spatially resolve scales considerably smaller than the typical scale-lengths of any structures in the observed object.

As noted in \citet{2005ApJ...621..227M}, individual clouds are not spatially or spectrally resolved when using the Echelle Spectrograph and Imager on Keck II. To represent this, the algorithm creates a grid of rays intersecting the computational grid at different points. A single spectrum is calculated by averaging the spectral lines over all rays produced by this grid. The total spatial size of this grid is the effective instrumental spatial resolution. The depth, width, and shape of the spectrum can vary strongly on this spatial resolution. Because the filling factor of dense clouds is small, at very high effective instrumental spatial resolution the grid of rays will generally either all pass through a single cloud, or entirely miss any cloud. This will produce either a narrow and saturated absorption line, or the complete absence of any absorption lines. At coarser effective instrumental spatial resolutions the line strengths can be weak (as is observed), because only a small fraction of rays will intersect clouds. Furthermore, at coarse instrumental resolution a large number of clouds are included within the simulated aperture, causing non-thermal broadening in the sodium-line, which is the effect we hope to capture.

\begin{figure}
\begin{center}
\includegraphics[width=.49\columnwidth]{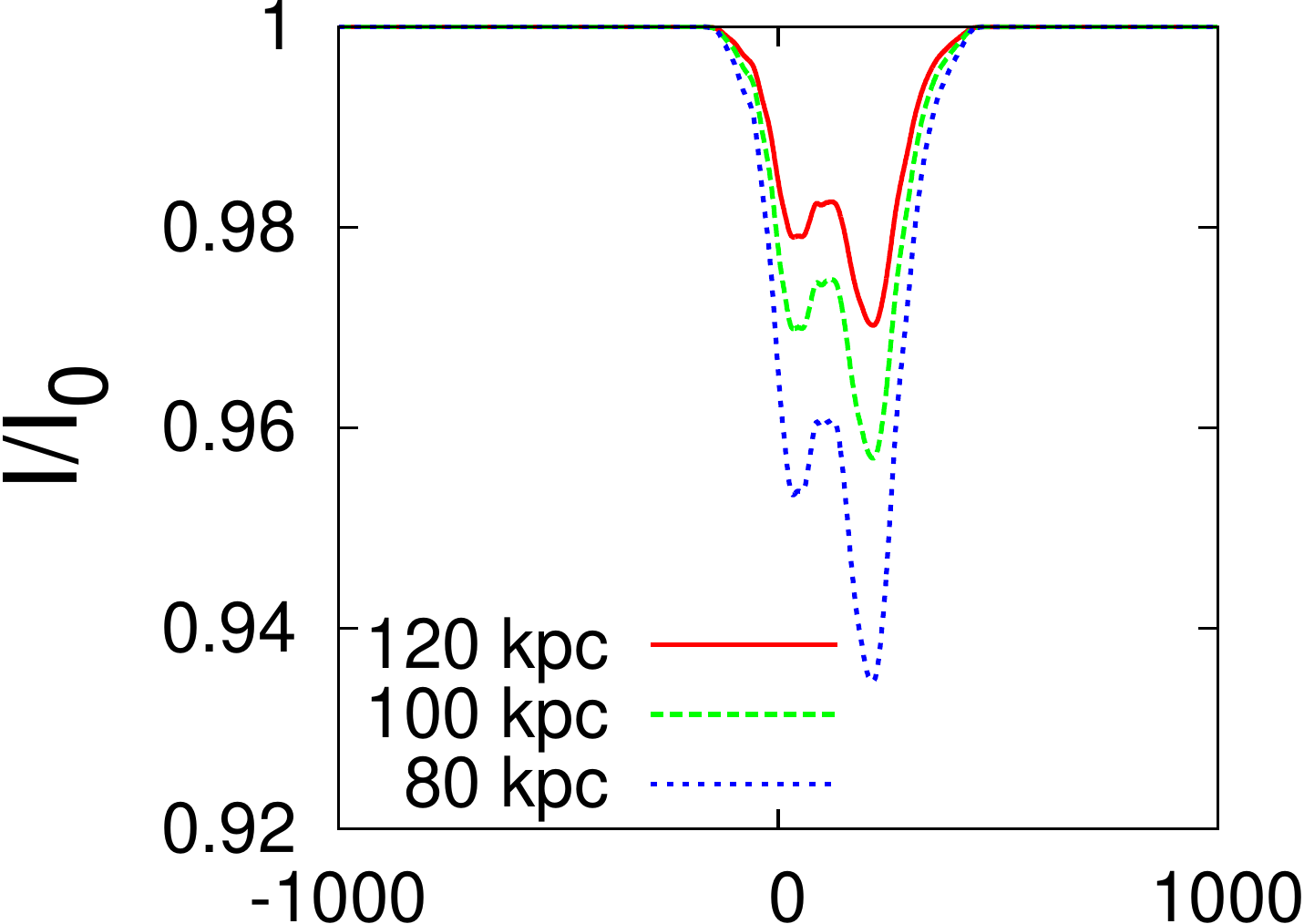}~\includegraphics[width=.49\columnwidth]{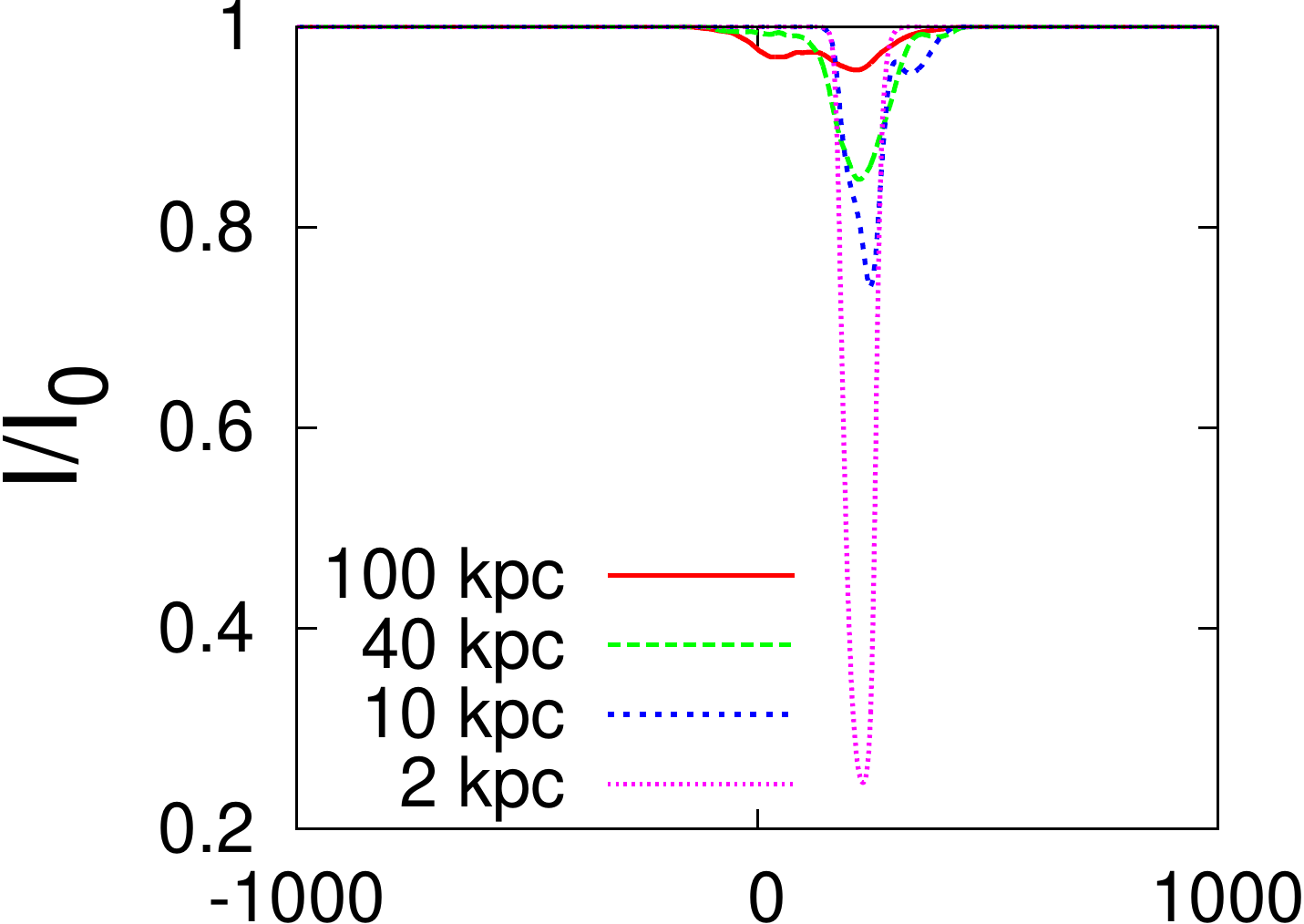}\\
\includegraphics[width=.49\columnwidth]{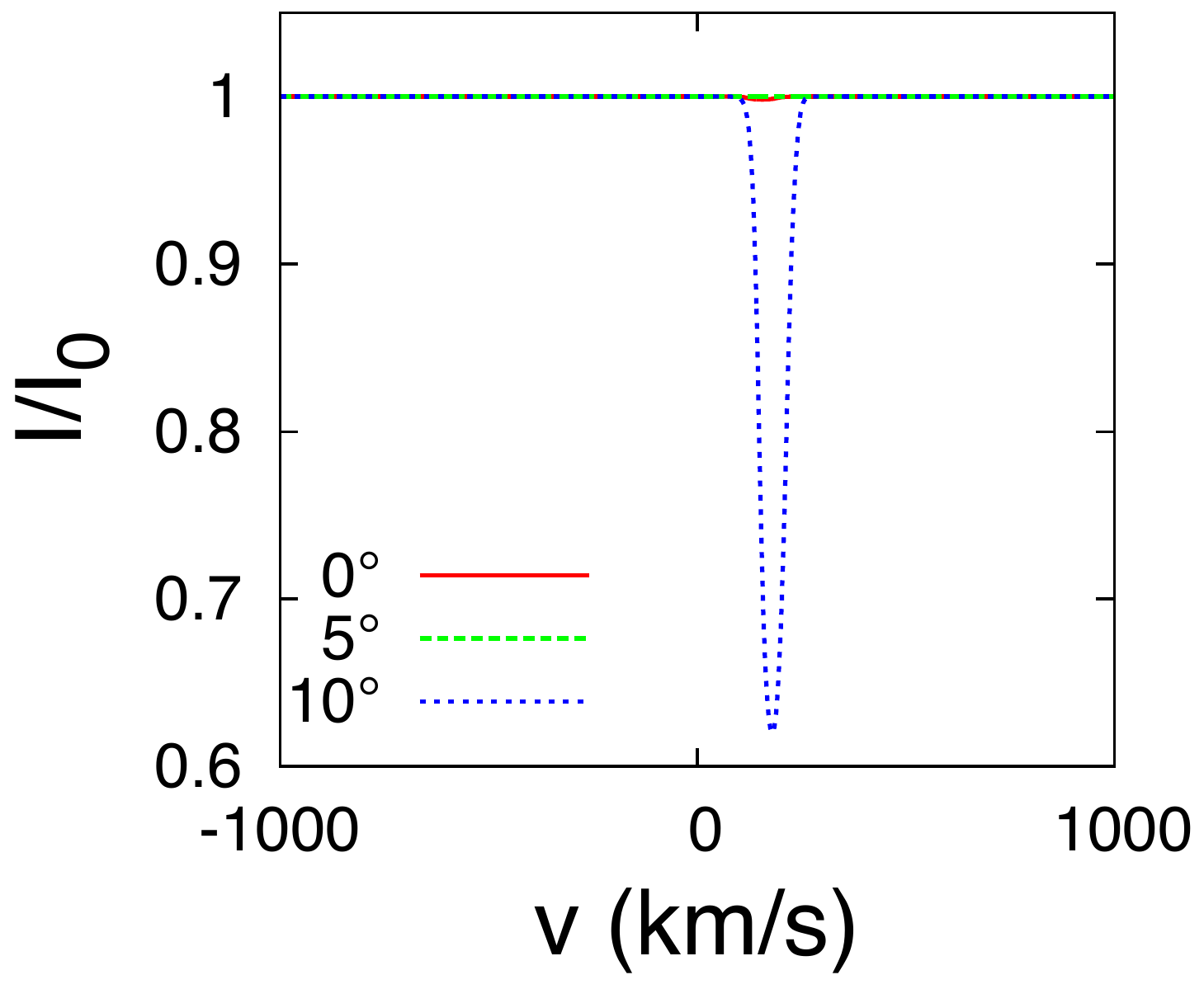}~\includegraphics[width=.49\columnwidth]{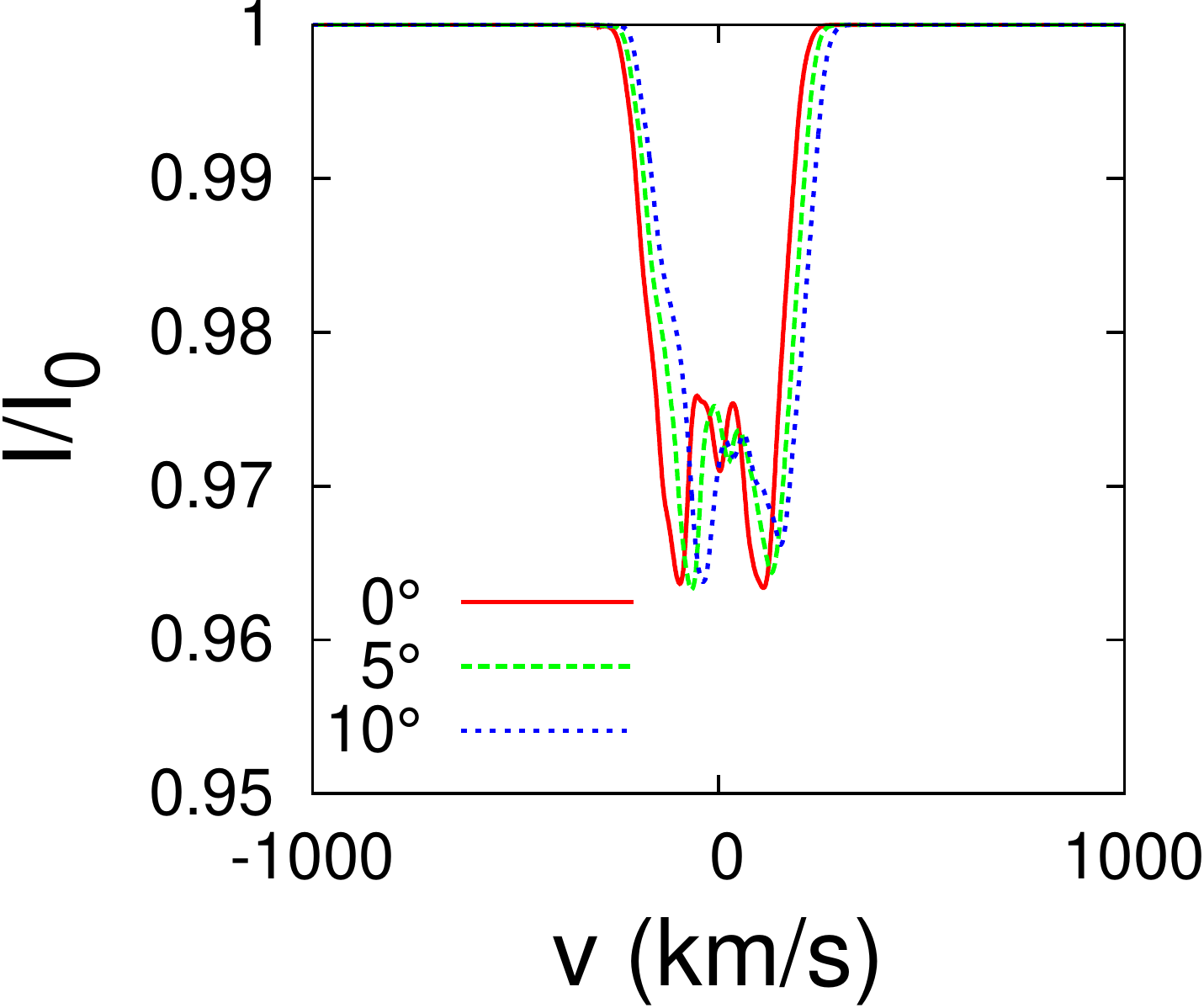}\\
\end{center}
\caption{\label{spatialresray}
Examining the effect of spatial resolution on line profiles. Top left: Lines with spatial resolutions of $80-120$ kpc. Top right: Lines with spatial resolutions of $2-100$ kpc. Bottom left: $2$ kpc spatial resolution line at three different elevation angles. Bottom right: $100$ kpc spatial resolution line at three different elevation angles.
}
\end{figure}

The instrumental resolution in \citet{2005ApJ...621..227M} is given as $\approx1^{\prime\prime}$, which corresponds to a spatial resolution of $\approx1$ kpc at a distance of $z\approx1$. For the galaxy centre models, this is larger than the entire simulation domain, and so the best approximation is to tighten the effective instrumental spatial resolution to be equal to the size of the entire simulation domain, i.e. $200$ pc. In the full galaxy models, the smallest cells are $250$ pc in size, and an instrumental resolution of $2$ kpc is equivalent to only $\approx8$ cell-widths. While this may be sufficiently large to resolve the entirety of a small cloud, it is not sufficient to resolve a large number of clouds and produce the observed non-thermal broadening. However, based on the assumption that the filling factor and velocity dispersion of dense NaI-absorbing clouds do not depend strongly on scale, we should expect the absorption spectrum of a real galaxy to not strongly depend on instrumental spatial resolution, and the effective instrumental spatial resolution in the simulated spectra can justifiably be coarsened from $\approx1$ kpc to a much larger value, and hence a significant number of clouds will be represented in the spectra. The assumption of scale-independence is not unreasonable, as these clouds are produced by turbulent motions, and the statistics of turbulence do not depend strongly on scale.

We have plotted the effects of varying the effective instrumental spatial resolution on the line profile of BHighRes in Fig.~\ref{spatialresray}. For the range $80-120$ kpc, all of the NaI absorbing gas is within the beam. At these coarse instrumental spatial resolutions, changing the resolution does not change the line profile shape, but only changes the strength of the line, as more rays ``miss'' the NaI absorbing regions entirely. The line-depth here is proportional to $1/l^2$ where $l$ is the instrumental spatial resolution. As $l$ drops further (e.g. $40$ kpc and below), the line not only strengthens, but noticeably changes its shape. However, this is strongly dependent on the ray's path --- as noted above, for narrow beams the ray generally either hits one cloud, or misses every cloud. This sensitivity is not the case for coarser spatial resolutions, and so $100$ kpc is chosen as the effective instrumental spatial resolution. The strength of the line is still strongly dependent on instrumental resolution in this regime, and so it is not meaningful to compare this line-strength to observations. However, the velocity-width and velocity at line centre should not be affected by these issues (given the assumption of only weak scale-dependence), and these are the properties that we emphasize for comparison with observations.

The circular component of velocity was also removed before calculating the line profiles, as at this very coarse spatial resolution circular motion can be a non-turbulent source of line-broadening that is not present in the observations due to their finer resolution. At later times, much of the rotation is no longer coherent, but some residual rotational velocity remains. To remove this, the mass and horizontal angular momentum of each FLASH cell are binned into annuli $2$ kpc thick and $2$ kpc tall to produce a set of rotation curves. The rotational velocity of each cell is then calculated with a 2-parameter linear interpolation from this table and subtracted from the cell.

The shape of the spectrum depended weakly on the number of rays produced by the grid - i.e. the numerical resolution of the raytracing. For a small number of rays (e.g. $32\times32$), the produced spectrum was jagged and did not show the smooth shape produced when a larger number of rays (e.g. $512\times512$) were used. The line was fairly well converged for an $L\times L$ grid for $L>64$. We hence used a $128\times128$ grid of rays as it was slightly more accurate and still computationally efficient.

To represent finite {\em spectral} resolution, the output line profile is convolved with a gaussian profile. Explicitly, we define
\begin{equation}
I_\mathrm{smooth}(v) = \int_{v_\mathrm{min}}^{v_\mathrm{max}} I(v^\prime) \frac{1}{\sigma \sqrt{\pi}} \exp \left(\frac{(v-v^\prime)^2}{\sigma^2}\right) dv^\prime
\end{equation}
where $\sigma$ is the standard deviation of the spectral broadening. This is calculated this using the convolution theorem, making use of the {\em Fastest Fourier Transform in the West} (FFTW) library \citep{2012ascl.soft01015F}. The full-width at half-maximum ($\sigma=\mathrm{FWHM}/(2\sqrt{2\ln{2}})$) of the Gaussian is set to $65$ km/s as in \citet{2009ApJ...698..693F}, to match the observations of \citet{2005ApJ...621..227M}.

\subsubsection{Line fitting}

Having produced line profiles over a large range of angles for a series of parameters, we fit a Gaussian to the result to estimate the line-width for comparison with observations. This was performed with a weighted $\chi^2$ fit. To ensure the line-centres were well-fit, the weight of wavelengths with zero NaI opacity (i.e. $I=1$) was reduced by weighting each point in the line according to $W=(1.0001-I)^{2}$, where the value of $1.0001$ is used instead of $1.0$ prevent divide-by-zero errors. This improved the algorithm's ability to correctly fit the width and strength of each line.

In edge-on orientations, a strong double-peak is visible (as in the top-left panel of Fig.~\ref{width_angle1}). These two peaks have a similar speed relative to the rest frame of the galaxy (i.e. the peaks at are $\pm v_0$ for some $v_0$) and are clearly distinct in the line profile. However, the clumpy nature of the outflow can produce several peaks in the line that do not correspond to distinct outflow components --- they are merely part of the velocity dispersion {\em within} an outflow, and should be modelled by a single broad Gaussian. To automatically capture both cases, two fits are performed on each line --- one with a single Gaussian curve, and one with two Gaussian curves, both of whose parameters are allowed to vary freely. To determine which fit is most appropriate, we examine the fit parameters. If the two Gaussians have a similar velocity at line centre --- where ``similar'' is defined by $||v_1|-|v_2||/(|v_1|+|v_2|)<0.5$ --- then the double-peak fit is used. Otherwise, the single-line fit is used. In the double-peak case, the parameters of each peak (FWHM, line-centre velocity, line-strength) are averaged to produce one set of parameters for each spectrum.

\section{Cloud finding and tracking}\label{cloudfind}

To determine the properties of the cold clouds of NaI-absorbing gas produced in these simulations, we developed an algorithm to identify, track, and plot them. The basic assumption of this algorithm is that all cells of sufficiently low temperature or high density are cold cloud cells, and that any cell that is orthogonally adjacent to another cell is labelled as part of the same cloud. The temperature and density cuts are $T<5\times10^4$ K and $\rho>10^{-21} \mathrm{g/cm}^3$, the same criteria used for NaI-absorbing gas in the raytracing algorithm.

To track clouds between checkpoint files we use a method similar to \citet{taskertan}. For each cloud, the centre-of-mass position and velocity were calculated. These were used to predict the position of the cloud in the following checkpoint file, i.e. if the positions and velocities in successive files $\bm{v}_i,\bm{r}_i$ and $\bm{v}_{i+1},\bm{r}_{i+1}$ are separated by a time $\Delta t$, the estimated position is $\bm{r}_\mathrm{est}=\bm{r}_i + \bm{v}_i \Delta t$.
The cloud closest to this position is identified as the ``same cloud''.

In the simulations in which clouds are no longer significantly accelerated by the outflow, we predict the trajectory of clouds using the ballistic approximation. The gravity field is produced from the same subroutine used by FLASH to produce a table of gravitational accelerations during each simulation's initialization. The initial position and velocity of each cloud is calculated from the FLASH output by the algorithm detailed above. The cloud's position and velocity are integrated using the leapfrog algorithm with a fixed time-step. The numerical integration allows us to calculate the full vector components of the velocity and position.

\begin{figure}
\begin{center}
\includegraphics[width=.496\columnwidth]{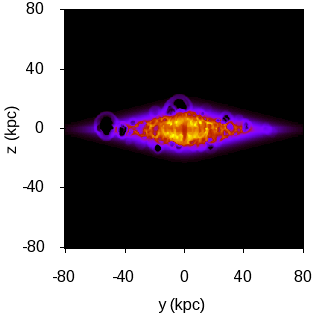}~
\includegraphics[width=.504\columnwidth]{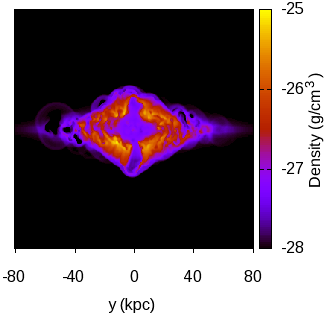}\\
\end{center}
\caption{\label{bigevolve}
Edge-on density slices of BHighRes at $t=20$ Myr (left) and $t=40$ Myr (right).
}
\end{figure}

\begin{figure}
\begin{center}
\includegraphics[width=.447\columnwidth]{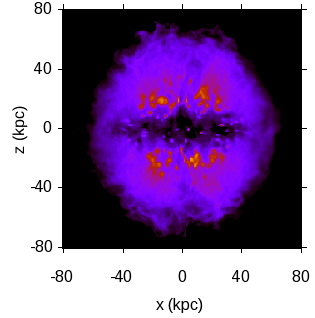}~
\includegraphics[width=.553\columnwidth]{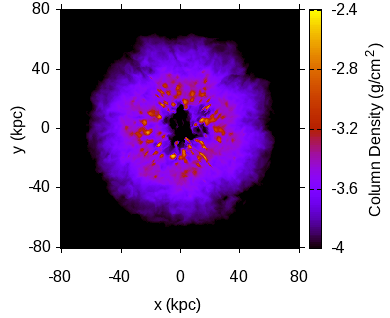}\\
\end{center}
\caption{\label{coldens0153}
Edge-on (left), and face-on (right) column density ray-trace plots of BHighRes at $t=100$ Myr.
}
\end{figure}

\section{Results}\label{resultscoldflows}

\subsection{Full Galaxy Models}\label{fullgalaxyresults}

\subsubsection{General Evolution}

In all models, the bubbles very quickly combine to form a single coherent outflow (see Fig.~\ref{bigevolve}). Although feedback is spread throughout the disc, and all of the disc gas develops a large vertical velocity, the flow is strongest near the centre, where star formation is most rapid. As almost all of the disc gas is incorporated into the outflow, the outflow front is very dense. This cool, dense front is followed by a hot low-density medium which has been directly heated by feedback. In some of the simulations, cool clouds are carried with the wind, while in others the interior of the wind remains hot. Column density plots of BHighRes in Fig.~\ref{coldens0153} illustrate these clouds. Their origin does not appear to be the Rayleigh-Taylor instability acting on the cold front of the super-bubble, as we detail in the following subsection.

\subsubsection{Formation and evolution of clouds}

\begin{figure}
\begin{center}
\includegraphics[width=.95\columnwidth]{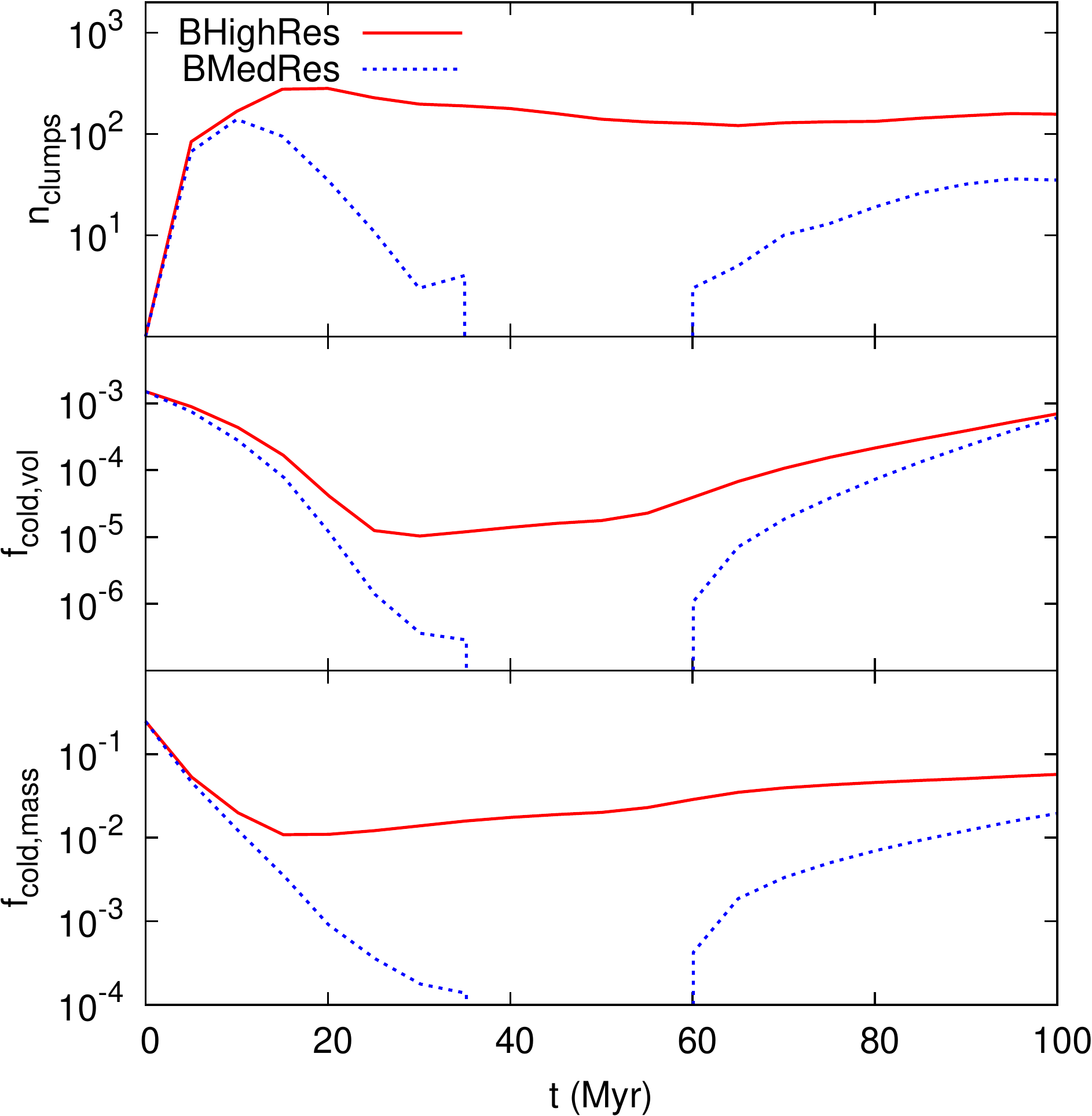}
\end{center}
\caption{\label{nclumpsbig}
Top: Number of clouds in full galaxy models. Centre: Cold ($T<5\times10^4$ K) gas volume fraction. Bottom: Cold gas mass fraction.
}
\end{figure}

\begin{figure*}
\begin{center}
\includegraphics[width=.435\columnwidth]{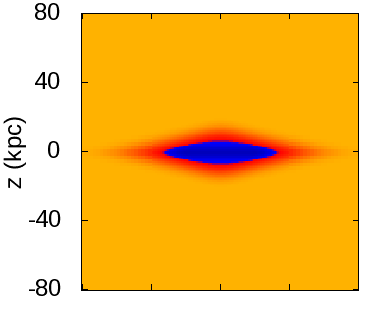}
\includegraphics[width=.359\columnwidth]{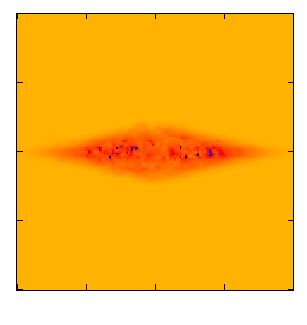}
\includegraphics[width=.359\columnwidth]{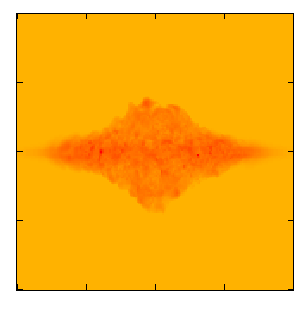}
\includegraphics[width=.359\columnwidth]{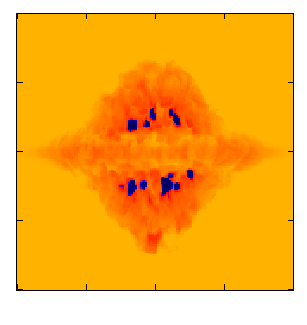}
\includegraphics[width=.488\columnwidth]{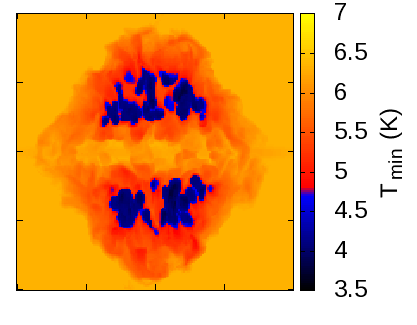}
\includegraphics[width=.435\columnwidth]{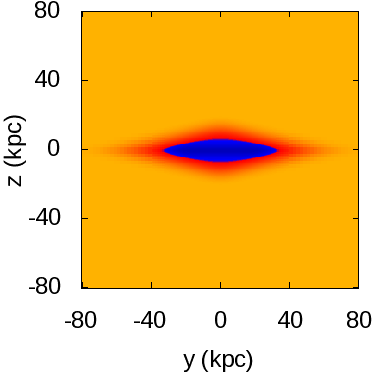}
\includegraphics[width=.359\columnwidth]{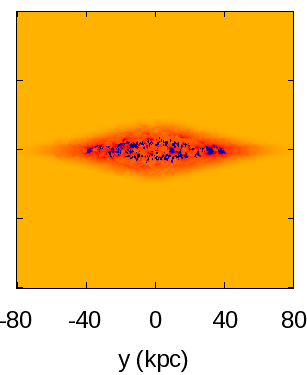}
\includegraphics[width=.359\columnwidth]{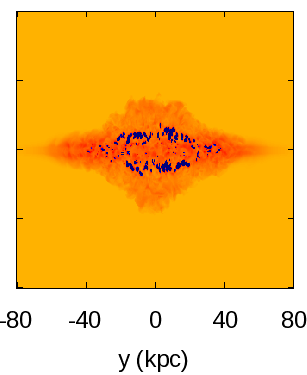}
\includegraphics[width=.359\columnwidth]{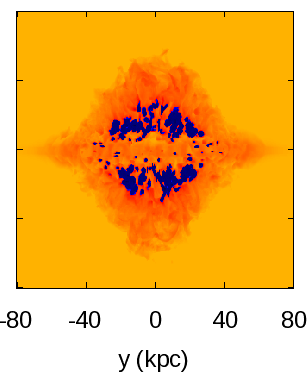}
\includegraphics[width=.488\columnwidth]{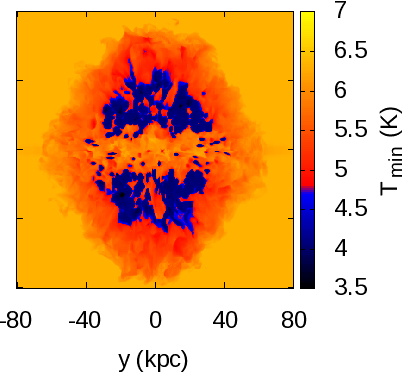}
\end{center}
\caption{\label{coldflattened}
Evolution of cold gas in BMedRes (top), and BHighRes (bottom), in snapshots at $t=0,25,50,75,$ and $100$ Myr (left to right). These plots show the temperature of the coldest gas of all cells along the line of sight (the $x$ direction), and hence the proximity of clouds is exaggerated. The colour scheme has also been chosen to distinguish the $T<5\times10^4$ K gas which will produce NaI absorption.
}
\end{figure*}

The evolution of the number of clouds and the gas gas mass and volume fractions in these models is plotted in Fig.~\ref{nclumpsbig}. BMedRes shows an initial peak in cloud formation, which is extinguished within $40$ Myr. BMedRes then forms additional clouds at $60$ Myr, and while BHighRes also shows an initial peak, it manages to maintain a large number of clouds throughout the simulation. The reason for this is apparent when we track the evolution of this cold gas, as shown in Fig.~\ref{coldflattened}. The disc initially cools, producing a large quantity of absorbing gas. Feedback bubbles divide this gas into discrete regions, which are identified as separate clouds of cool gas. In all models except BHighRes, the feedback fills the entire disc with hot gas, destroying all of the cold regions. However, in BHighRes, the cold clouds are not destroyed, and instead are pushed out of the disc by the hot winds, forming the cold high-velocity-dispersion component of the wind while continuing to accrete cooling gas from the surrounding hot wind. This is confirmed by the centre and bottom panels of Fig.~\ref{nclumpsbig}, where the volume and mass fractions of cold gas in BMedRes dramatically drop by $t=40$ Myr, while the cold gas in BHighRes persists more effectively. These cool regions can only form above and below the plane of the disc, as the feedback is too intense in the plane to allow any cold gas to exist. In BMedRes there is also a cold high-velocity dispersion component in the wind, but this is caused solely by gas cooling within the wind. This cooling happens because the starburst has completed, and because the hot gas is now free to adiabatically expand into the regions above the disc.

Hence --- unlike the findings of \citet{2009ApJ...698..693F} --- the cold gas in BHighRes is produced by the cool regions between hot bubbles being pushed out of the disc by the pressure of the hot outflow beneath them, while the cold gas in both BMedRes and further cold gas in BHighRes is produced by cooling in the turbulent outflow itself \citep[as in][]{2013ApJ...763L..31S}.

\begin{figure}
\begin{center}
\includegraphics[width=.98\columnwidth]{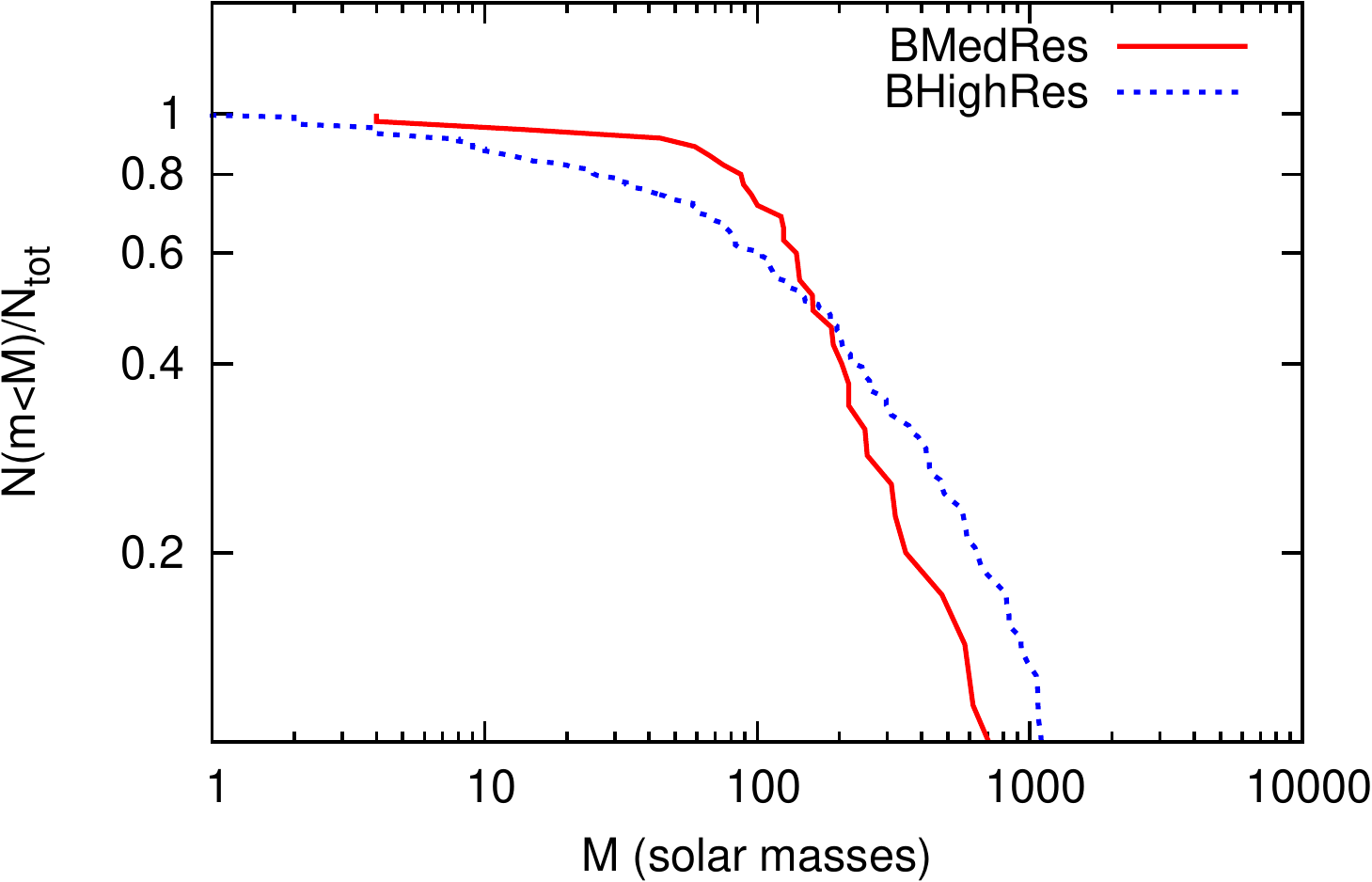}\\
\end{center}
\caption{\label{bigmassspect}
Cumulative mass spectra for BHighRes and BMedRes.
}
\end{figure}

The formation of cold clouds of gas in these simulations is resolution-dependent, and so even at our highest resolution model (BHighRes) only a moderate number ($\sim150$) of these clouds form by the end of the simulation time, as visible in a ray-traced column-density plot (Fig.~\ref{coldens0153}). Only one of these clouds lies in the $x=0$ plane, which suggests that the additional avenues for gas flow permitted in 3D simulations inhibit their formation. The number and mass of these clouds depends on resolution, with fewer clouds forming at lower resolution. As the mass spectrum plots in Fig.~\ref{bigmassspect} demonstrate, the clouds at higher resolution are also more massive, possibly because we can resolve higher densities, and hence cooling instabilities are more dramatic, but also because these clouds have formed at an earlier time from a denser medium (i.e. the disc rather than the outflow).

\begin{figure}
\begin{center}
\includegraphics[width=.48\columnwidth]{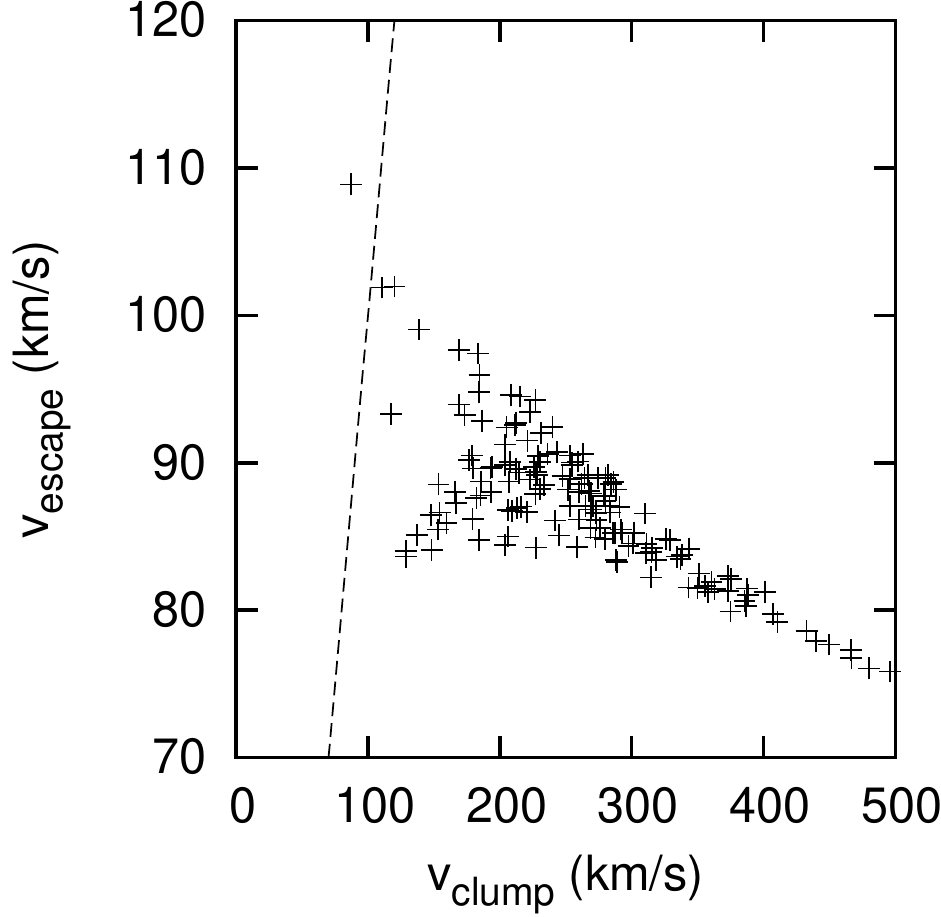}~\includegraphics[width=.48\columnwidth]{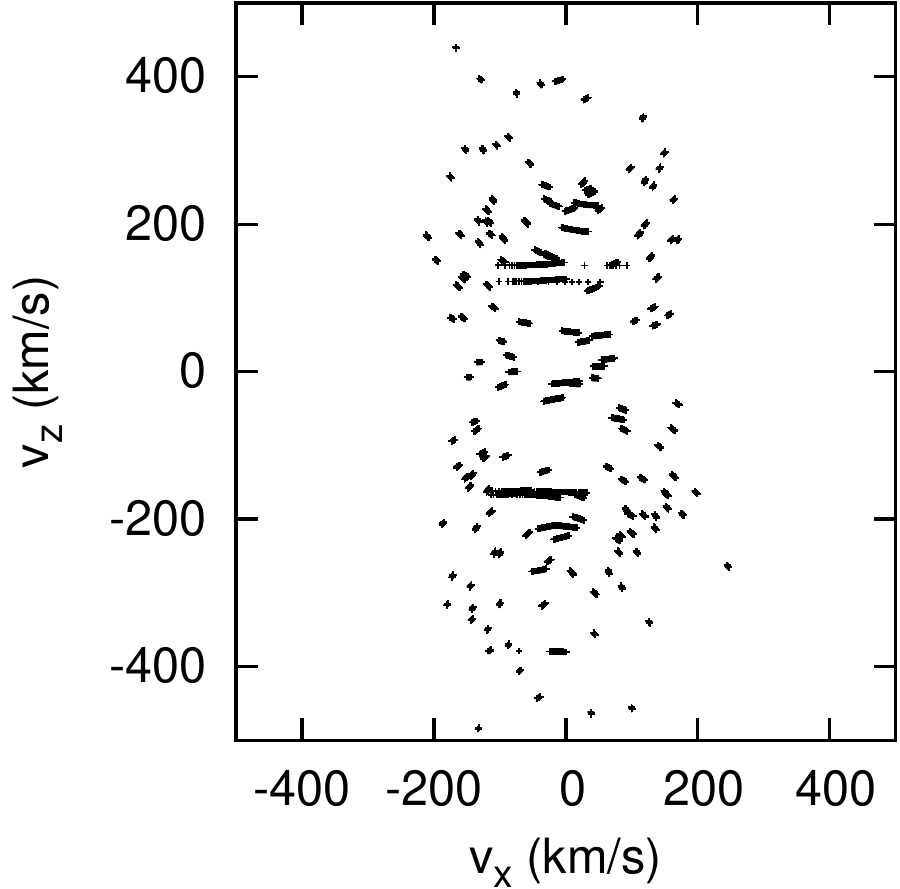}\\
\end{center}
\caption{\label{vtraj0153}
Left: Escape velocities and net velocities for all clouds. The line indicates where $v_\mathrm{escape}=v_{cloud}$.
Right: Velocity space trajectories for all clouds. The clouds do not decelerate significantly over $100$ Myr.
}
\end{figure}

The escape velocities of these clouds are calculated using $v_\mathrm{escape}=\sqrt{2\phi}$, where $\phi$ is the gravitational potential given by FLASH's subroutines --- i.e. $d\bm{v}/dt = \nabla\phi$. Most of the clouds have exceeded the escape velocities at their positions (Fig.~\ref{vtraj0153}). Making use of the ballistic approximation detailed above, there is very little change in each cloud's velocity over a period of $100$ Myr, as also shown in Fig.~\ref{vtraj0153}. Ignoring hydrodynamics, this would imply that most of the cold gas will not eventually rain back on to the disc as a ``galactic fountain'', but will instead continue outwards and enrich the intergalactic medium.

\begin{figure}
\begin{center}
\includegraphics[width=.5\columnwidth]{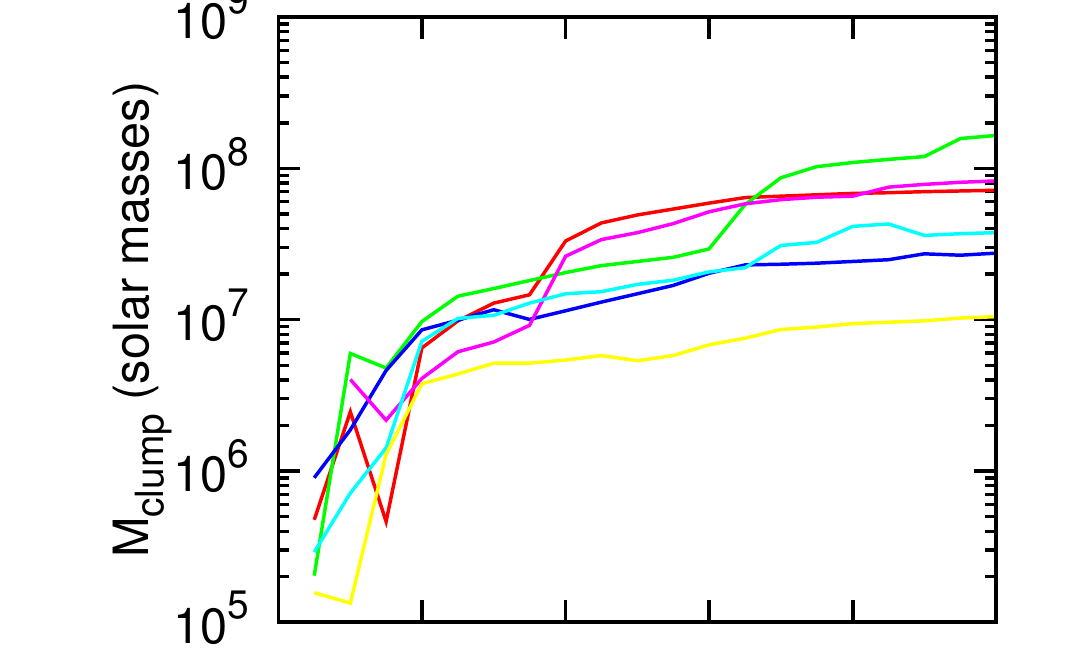}~\includegraphics[width=.5\columnwidth]{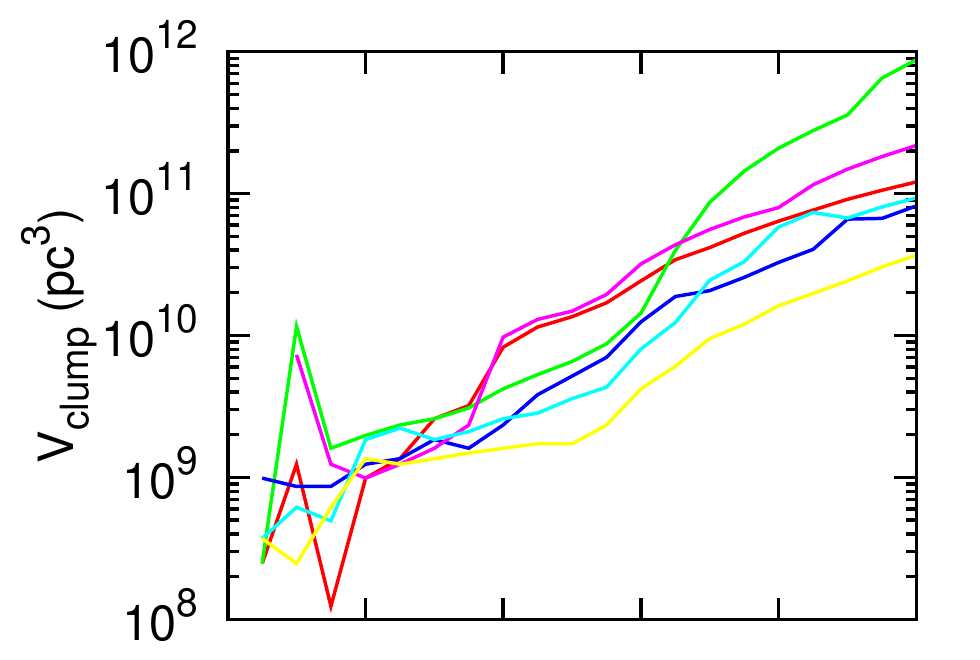}\\
\includegraphics[width=.5\columnwidth]{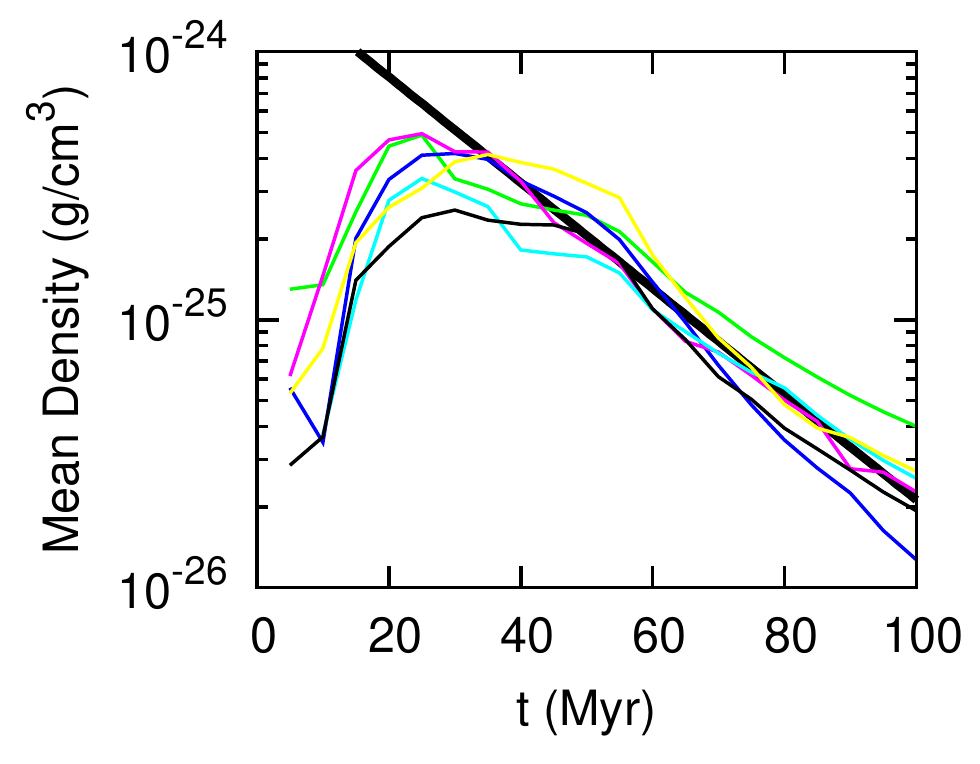}~\includegraphics[width=.5\columnwidth]{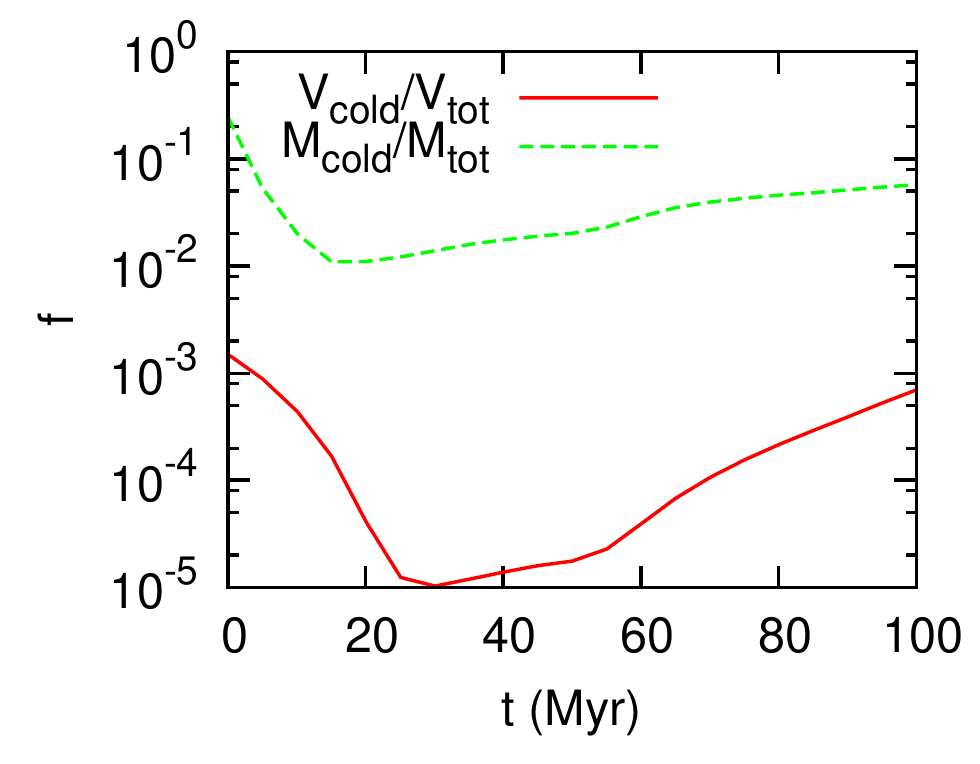}\\
\end{center}
\caption{\label{cold_track}
The histories of a sample of clouds in BHighRes. Top left: Mass history. Top right: Volume history. Bottom left: Mean density (i.e. mass/volume) history. The thick line is $2\times10^{-24}\exp(-t/22 \mathrm{Myr})$ g/cm$^3$. For these three plots, the different colours of the thin lines are to distinguish between the different clouds in the sample. Bottom right: The evolution of the mass fraction and volume fraction of cold gas in BHighRes. 
}
\end{figure}

The next question is whether we might expect hydrodynamic effects to dissolve the clouds. This is estimated by examining their history to see if they have begun to lose mass. The mass history of a sample of clouds is plotted in Fig.~\ref{cold_track}, and it is clear that clouds are still {\em gaining} mass, as additional gas is cooling and being accreted. This fits the overall trend of cold gas also plotted in Fig.~\ref{cold_track}. The large density of gas ejected from the galaxy permits efficient cooling, catalysing cloud-forming instabilities.

However, even though the clouds are gaining mass, the mean density of each cloud is dropping exponentially (Fig.~\ref{cold_track}). If this continues, clouds will reach the background density of $10^{-29} \mathrm{g/cm}^3$ in only $\approx250$ Myr, and will likely be disrupted by hot flows well before then. This suggests that the clouds will not likely remain coherent as they rise to large distances from the disc. \citet{2006ApJ...647..222M} observed that NaI absorption extends out to distances of around $4$--$18$ kpc. Although the hot outflows in BHighRes extend much further than this, out to almost $100$ kpc, the cold clouds are closer to the disc, agreeing with this result. 

However, X-ray emission maps from Chandra surveys \citep{2003ApJ...592..782P} only reveal hot gas at scales of $\approx10$ kpc, much closer to the disc --- either the hot gas at large altitudes must be currently undetectable in X-ray wavelengths (perhaps it is not dense enough), or our outflows are too energetic. We propose two possible explanations for an overly energetic outflow in these simulations.

\begin{figure}
\begin{center}
\includegraphics[width=.98\columnwidth]{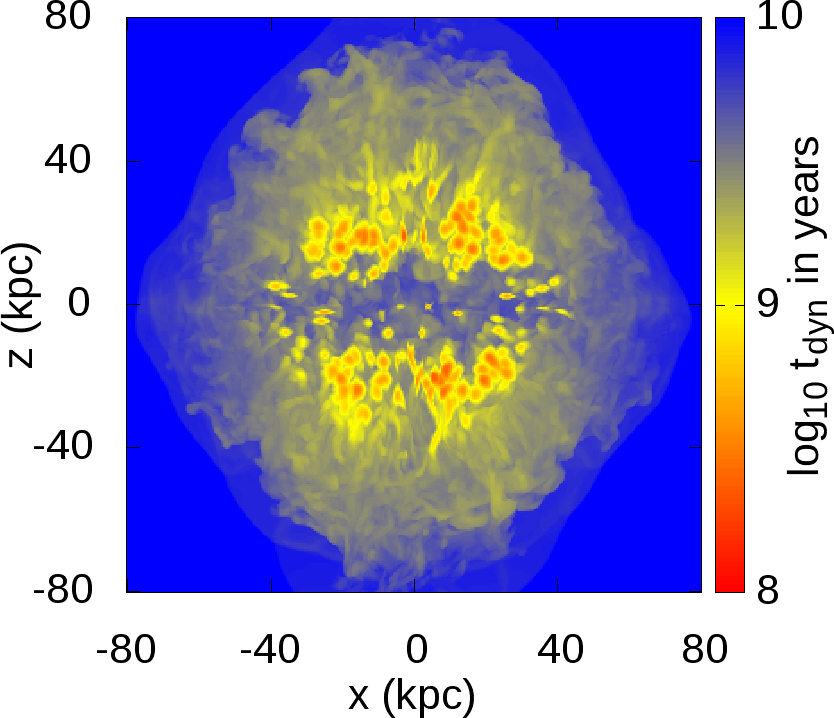}\\
\end{center}
\caption{\label{tdyn0153}
{The minimum dynamical time along the line-of-sight for an edge-on view of BHighRes at $t=100$ Myr.} 
}
\end{figure}

\begin{figure*}
\begin{center}
\includegraphics[width=.5224\columnwidth]{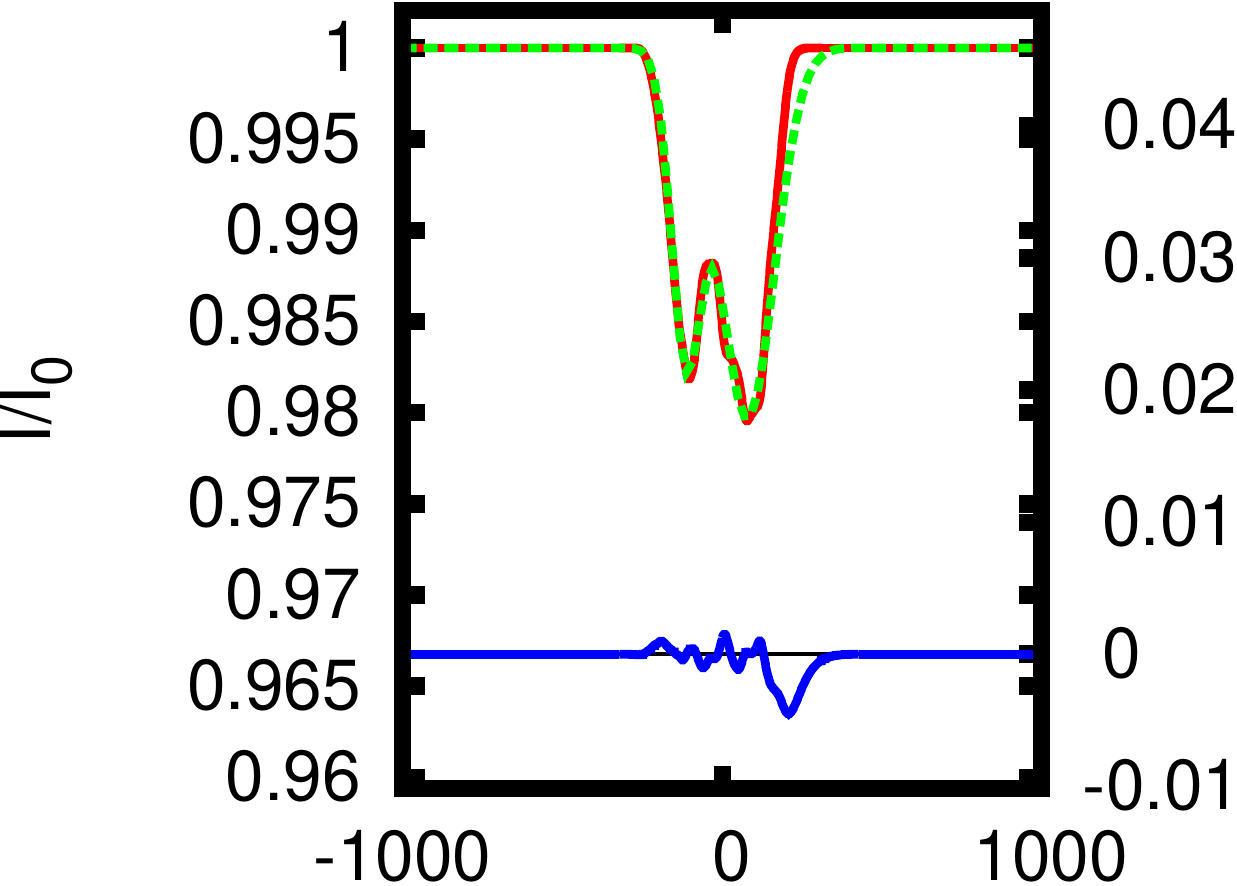}~\includegraphics[width=.4798\columnwidth]{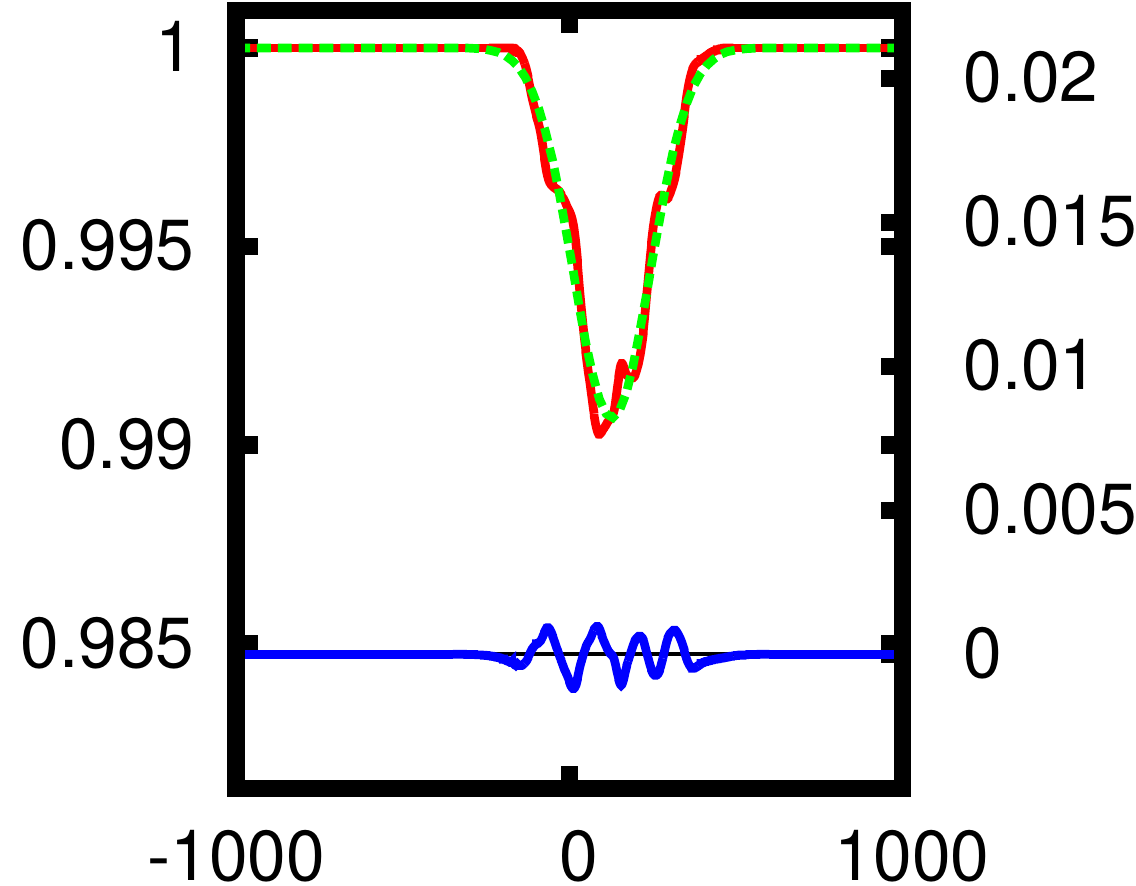}
\includegraphics[width=.4798\columnwidth]{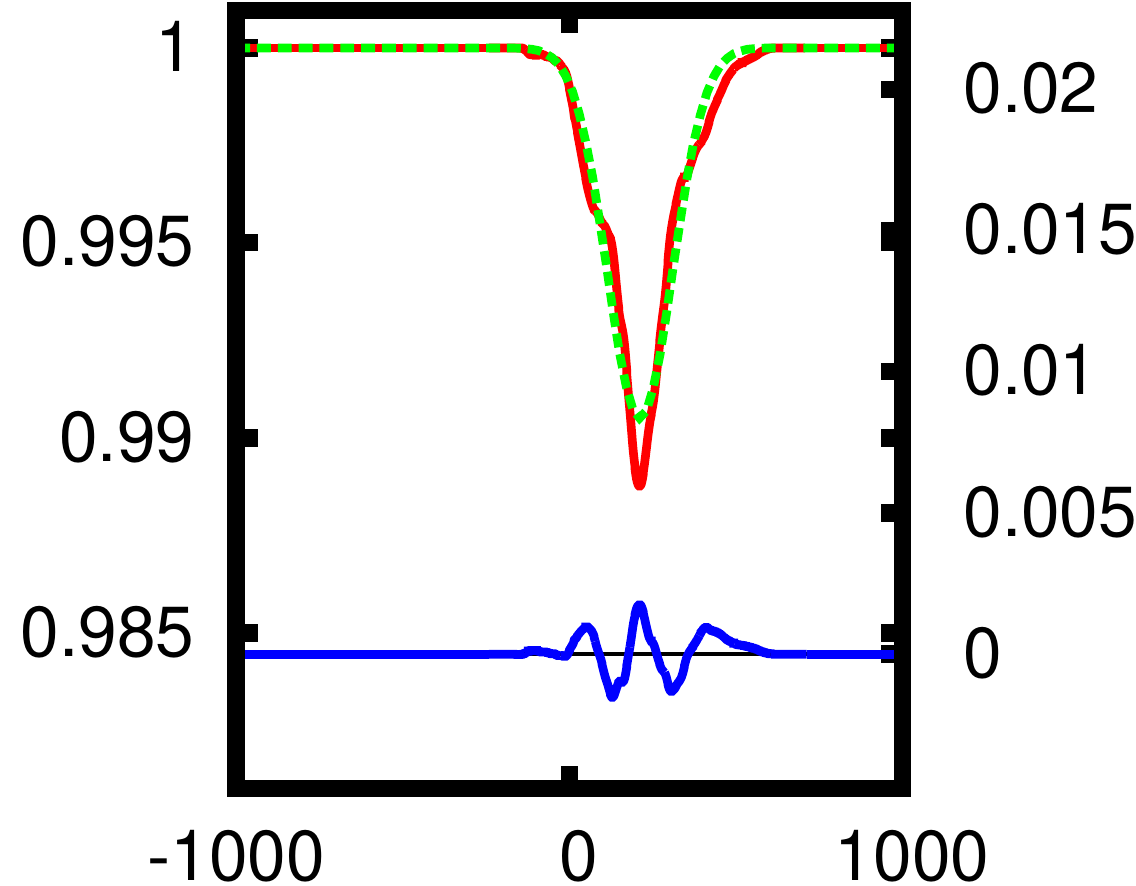}~\includegraphics[width=.5180\columnwidth]{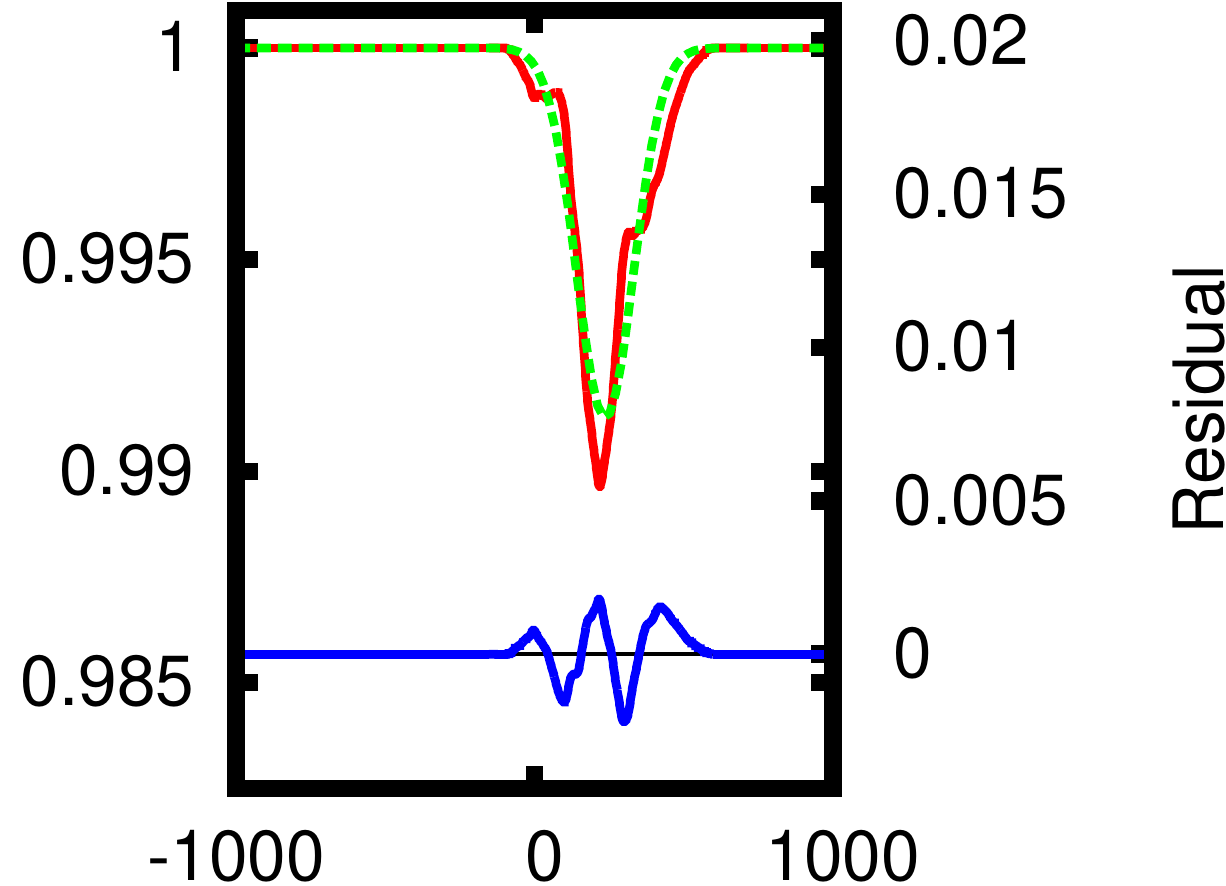}\\
\includegraphics[width=.5224\columnwidth]{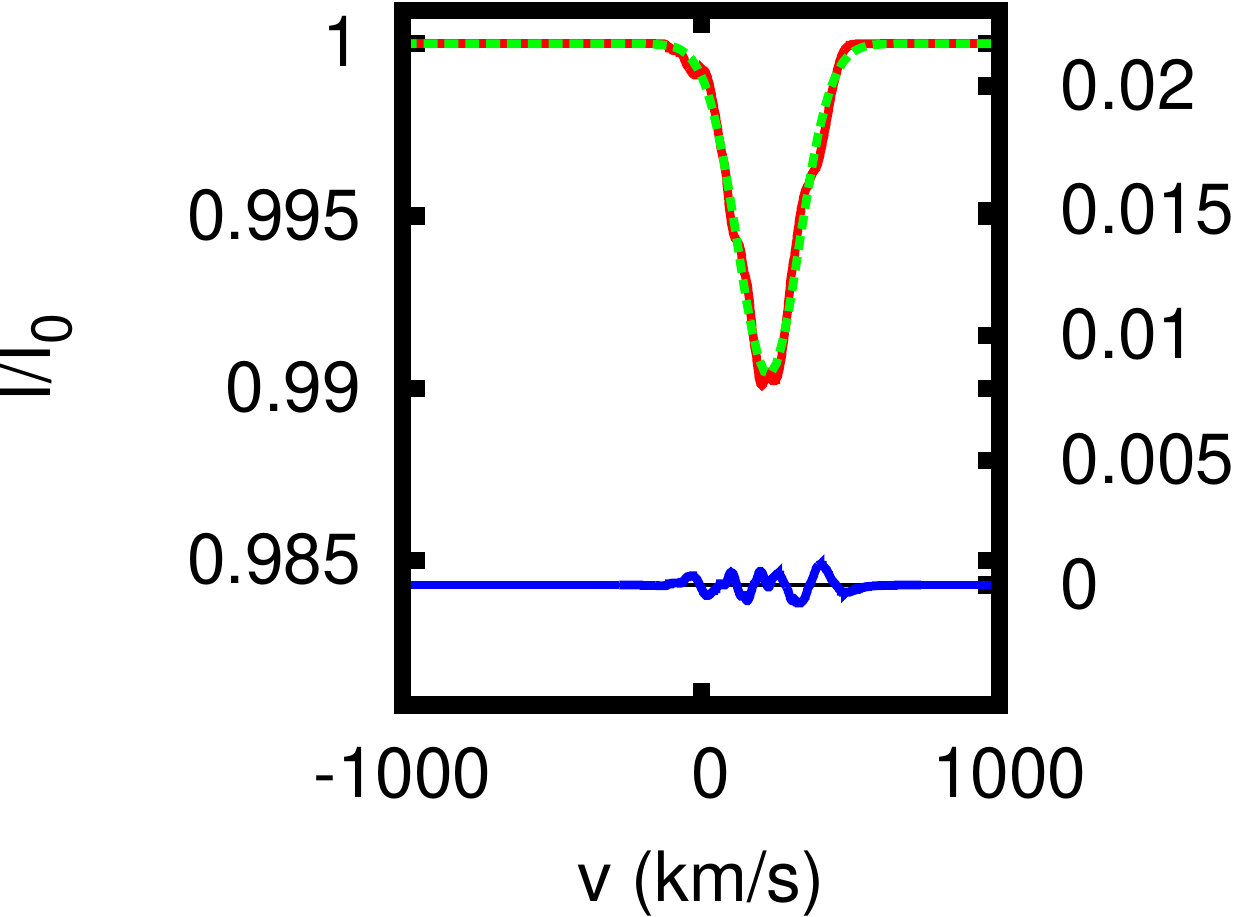}~\includegraphics[width=.4798\columnwidth]{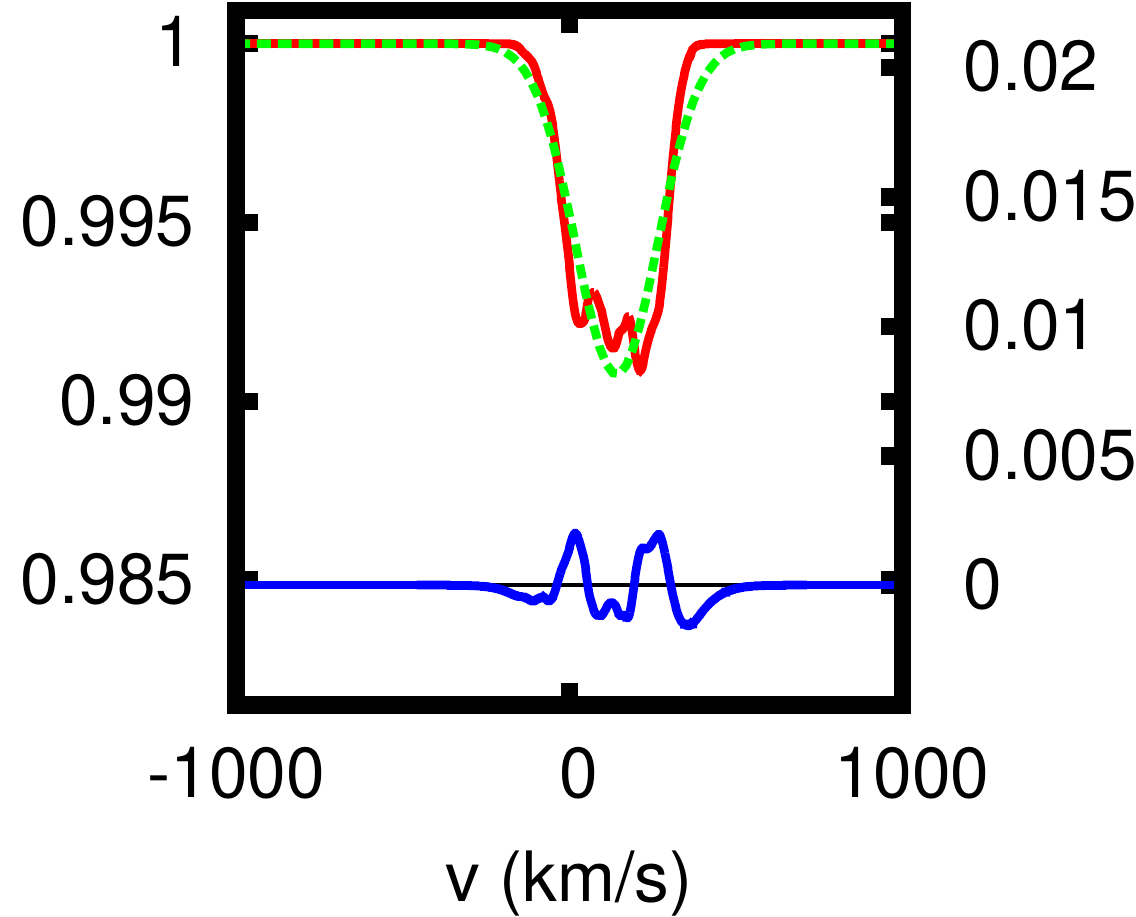}
\includegraphics[width=.4798\columnwidth]{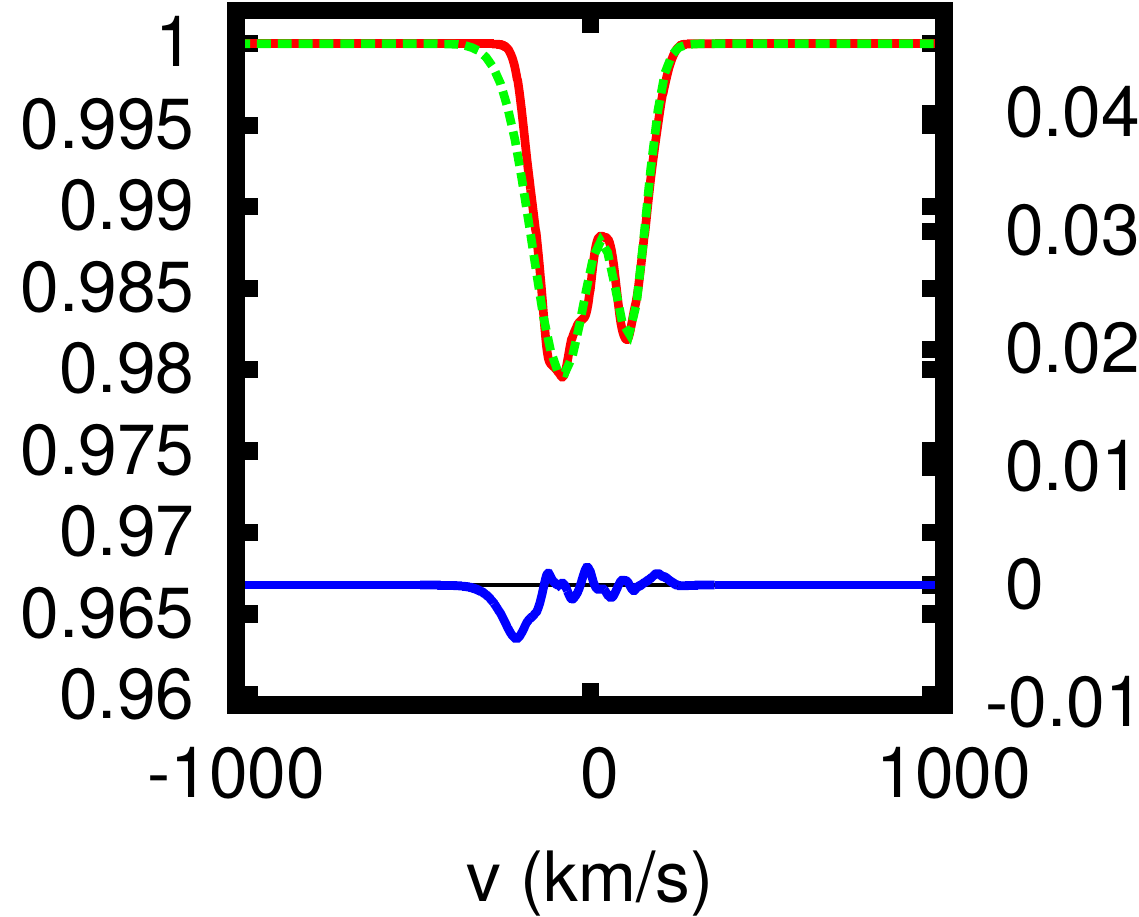}~\includegraphics[width=.5180\columnwidth]{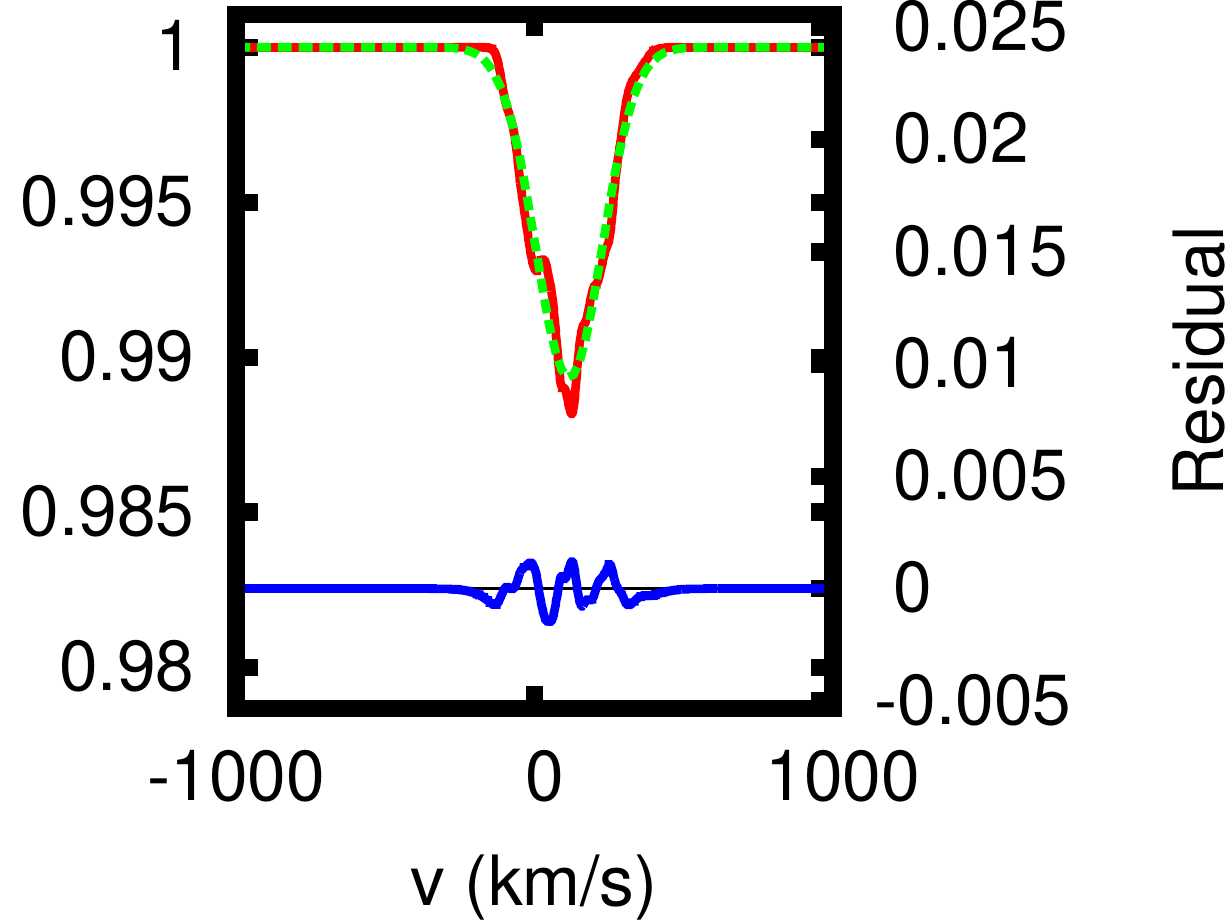}\\
\end{center}
\caption{\label{width_angle1}
Line profiles for elevation angles of $0-210^\circ$ (where an elevation angle of $0^\circ$ is viewing the disc edge-on), and azimuthal angles of $45^\circ$, for the simulation BHighRes. The elevation angle increases from top-left to bottom-right in $30^\circ$ steps. The red line is the line profile, while the green line is a Gaussian fit, both described by the axis labels on the left of each plot. The y-axis is normalised so that the continuum intensity is equal to 1. The blue line is the residual, described by the axis labels on the right of each plot.
}
\end{figure*}

Firstly, the limited resolution does not permit modelling of the detailed structure of the interaction between hot bubbles and cool gas in the disc, and hence instead of hot under-dense gas escaping through narrow avenues, almost the entire gaseous mass of the disc is propelled outwards, providing a large reservoir of momentum to plough through the halo. This might be exacerbated by the stochastic feedback mechanism, which is not entirely consistent with realistic star formation, especially as resolution limits force a lower limit on the size of hot bubbles. For example, a hot bubble is likely to be placed directly on top of an existing hot bubble, even though star formation is unlikely in such a hot low-density environment.

The second explanation is that the halo model lacks density contrasts, and that a more detailed model of the gaseous halo will would affect the evolution of hot outflowing gas. A galaxy in a cosmological context - especially a merging system such as a ULIRG - may not have a spherically symmetric hot halo whose density smoothly drops with radius. For example, the presence of inflows (both hot and cold) and other substructure in the gaseous halo could provide an impediment to hot outflowing gas, resulting in a lower outflow velocity and a smaller spatial extent.

\subsubsection{Estimates of neglected effects on clouds}\label{bigneglect}

Some physical processes have been neglected in our simulations, or suppressed due to the limited resolution. Here we estimate the effects of neglecting self-gravity and thermal conduction, as well as cloud destruction from hot flows below our resolution.

The importance of self-gravity can be determined by calculating the dynamical time $t_\mathrm{dyn}=1/\sqrt{G\rho}$ and comparing it to the typical cloud age. The minimum dynamical time along the line-of-sight for an edge-on view of BHighRes at $t=100$ Myr is plotted in Fig.~\ref{tdyn0153}. In most of the volume $t_\mathrm{dyn}>1$ Gyr. The minimum dynamical time in the entire domain is $t_\mathrm{dyn}=160$ Myr. As the clouds form at around $20$ Myr, the oldest clouds are $80$ Myr old. Hence self-gravity, while not insignificant, is not likely to be dominant in this model.

We can determine the importance of thermal conduction by comparing the conductive heating rate of a cloud with its radiative cooling rate. The cooling rate per unit volume is $\Lambda n_c^2$, where $n_c$ is the number-density of particles in the cold clouds, and $\Lambda$ is the cooling function at the appropriate temperature. In the classical limit \citep{1962pfig.book.....S,1990ApJ...358..375B}, the heating rate can be approximated by
\begin{equation}
\nabla\cdot q = -\nabla\cdot\kappa\nabla T_h \approx \kappa T/l^2,
\end{equation}
where $\nabla\cdot q$ is the cooling rate per unit volume, $\kappa$ is the thermal conductivity, $T$  is the temperature of the hot gas, and $l$ is the typical diameter of the clouds. The conductivity of the electrons in the hot gas is $\kappa=5.6\times10^{-7}T^{5/2}$ erg s$^{-1}$ K$^{-1}$ cm$^{-1}$ \citep{1984ApJ...281..690D,1990ApJ...358..375B}.  For a cloud of $T=4\times10^5$ K at our chosen metallicity of $Z=0.05Z_\odot$, our cooling function gives $\Lambda=2.8\times10^{-23}$ erg cm$^3$ s$^{-1}$. We can choose typical values for the other parameters, setting $T_h\approx10^7$ K, $l\approx3$ kpc, and $n_c\approx0.06$ cm$^{-3}$. This gives a heating rate of $\nabla\cdot q\approx1.8\times10^{-26}$ erg s$^{-1}$ cm$^{-3}$, and a cooling rate of $\Lambda n_c^2\approx1.0\times10^{-25}$ erg s$^{-1}$ cm$^{-3}$. Hence if the classical assumption holds, thermal conduction only has a small contribution. We note that our chosen metallicity is less than that of recent observations \citep{2008ApJ...674..172R}. A metallicity six times larger than that used in our simulation will result in a cooling rate of $1.3\times10^{-25}$ erg s$^{-1}$ cm$^{-3}$, which is similarly greater than the heating rate, and hence using a higher metallicity should not drastically change our results.

We must also determine whether the classical assumption is appropriate. If the mean free path for electrons is not small compared to the size of clouds, the classical assumption does not apply, and thermal conduction is saturated and it will be weaker than the classical estimate \citep{1977ApJ...211..135C}. The mean effective free path for the electrons in the hot gas is given by $\lambda_\kappa = 0.284 \phi_c (T_e/10^7 \mathrm{K}) n_e^{-1}$ pc, where $\phi_c$ is a constant of order one. Taking typical values of $T_e\approx10^7$ K and $n_e\approx7\times10^{-4}$ cm$^{-3}$ gives $\lambda_\kappa\approx 400$ pc. This is only about an order of magnitude smaller than the typical cloud size in these full galaxy simulations, and so thermal conduction may be even weaker than the classical estimate.

Finally, we use the ``cloud crushing'' timescale of \citet{1994ApJ...420..213K} to estimate the effect of unresolved hot flows destroying the clouds. This is
\begin{equation}
t_\mathrm{cc}=\left(\frac{\rho_c}{\rho_h}\right)^{1/2}\frac{l_c}{|v_h-v_c|},
\end{equation}
where $\rho_h$ and $v_h$ are the density and velocity of the hot wind, $\rho_c$ and $v_c$ are the density and velocity of the cold clouds, and $l_c$ is the typical diameter of the clouds. Taking typical values of $\rho_h\approx10^{-29}$ g cm$^{-3}$, $\rho_c\approx10^{-25}$ g cm$^{-3}$, $v_h\approx300$ km s$^{-1}$, $v_c\approx260$ km s$^{-1}$, and $l_c\approx3$ kpc gives $t_\mathrm{cc}\approx 5$ Gyr, which is much longer than our simulation time. Hence unresolved processes in the hot wind are unlikely to directly disrupt the clouds.

\subsubsection{Simulated spectra}\label{specnoturb}

In Fig.~\ref{width_angle1} we have plotted absorption lines for BHighRes at $t=100$ Myr at viewing angles with elevation angles from $0-210^\circ$, where an elevation angle of $0^\circ$ corresponds to an edge-on view of the disc. These plots are for a single component of the NaI doublet, and also do not include instrumental broadening. The disc lines are not visible as the disc does not retain any cold gas in this simulation.

\begin{figure}
\begin{center}
\includegraphics[width=1.\columnwidth]{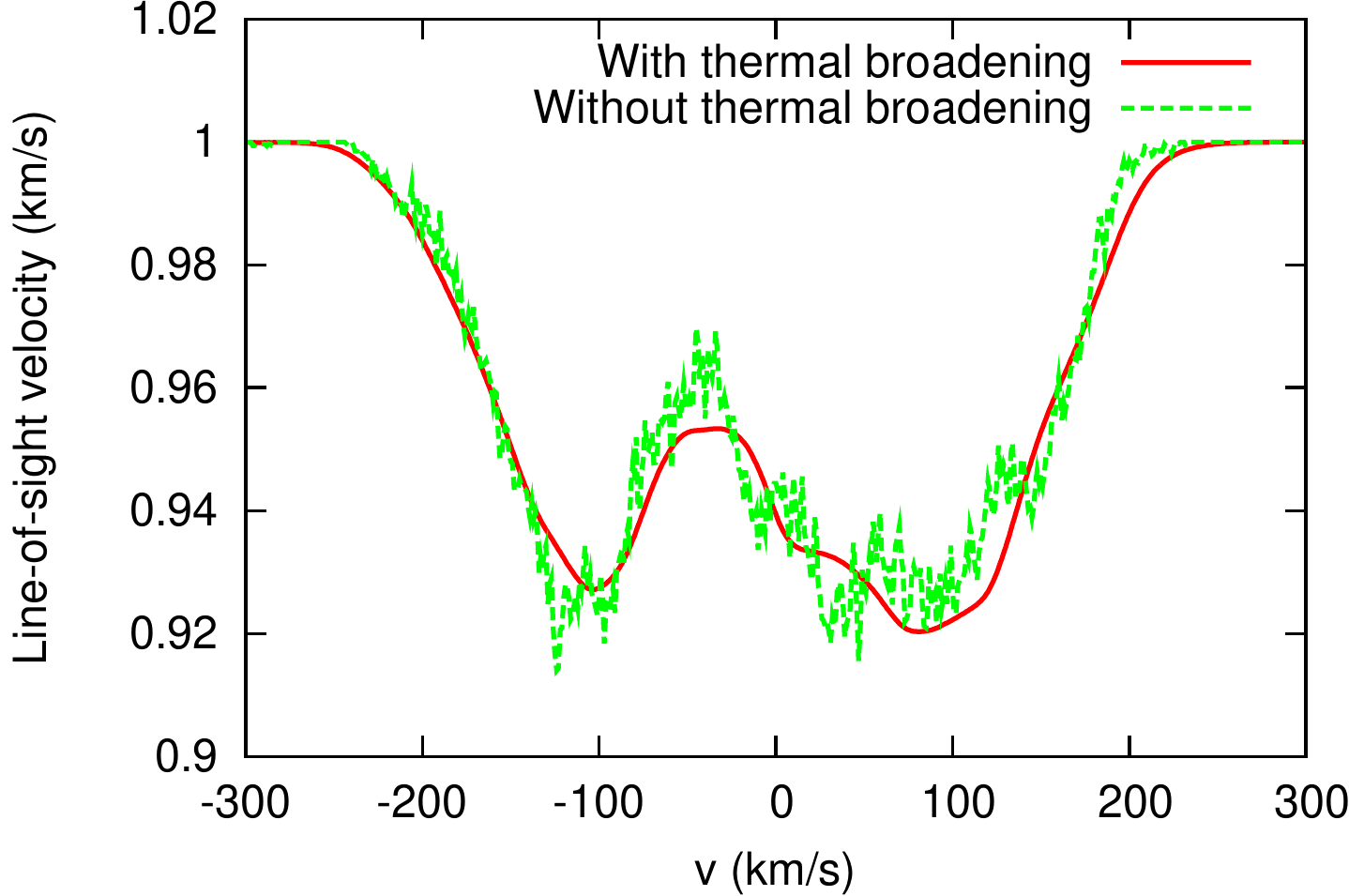}
\end{center}
\caption{\label{thermalnonthermal}
Line profiles with and without thermal (i.e. Doppler) broadening in a simulation. The strength of the un-broadened lines are dependent on the spectral resolution of the code, and so we arbitrarily rescale the intensity to allow a closer comparison with the thermal broadened lines. These lines are sight-lines through the BHighRes model at an elevation angle of $0^\circ$ (i.e. edge-on) and an azimuthal angle of $45^\circ$.
}
\end{figure}

\begin{figure}
\begin{center}
\includegraphics[width=.48\columnwidth]{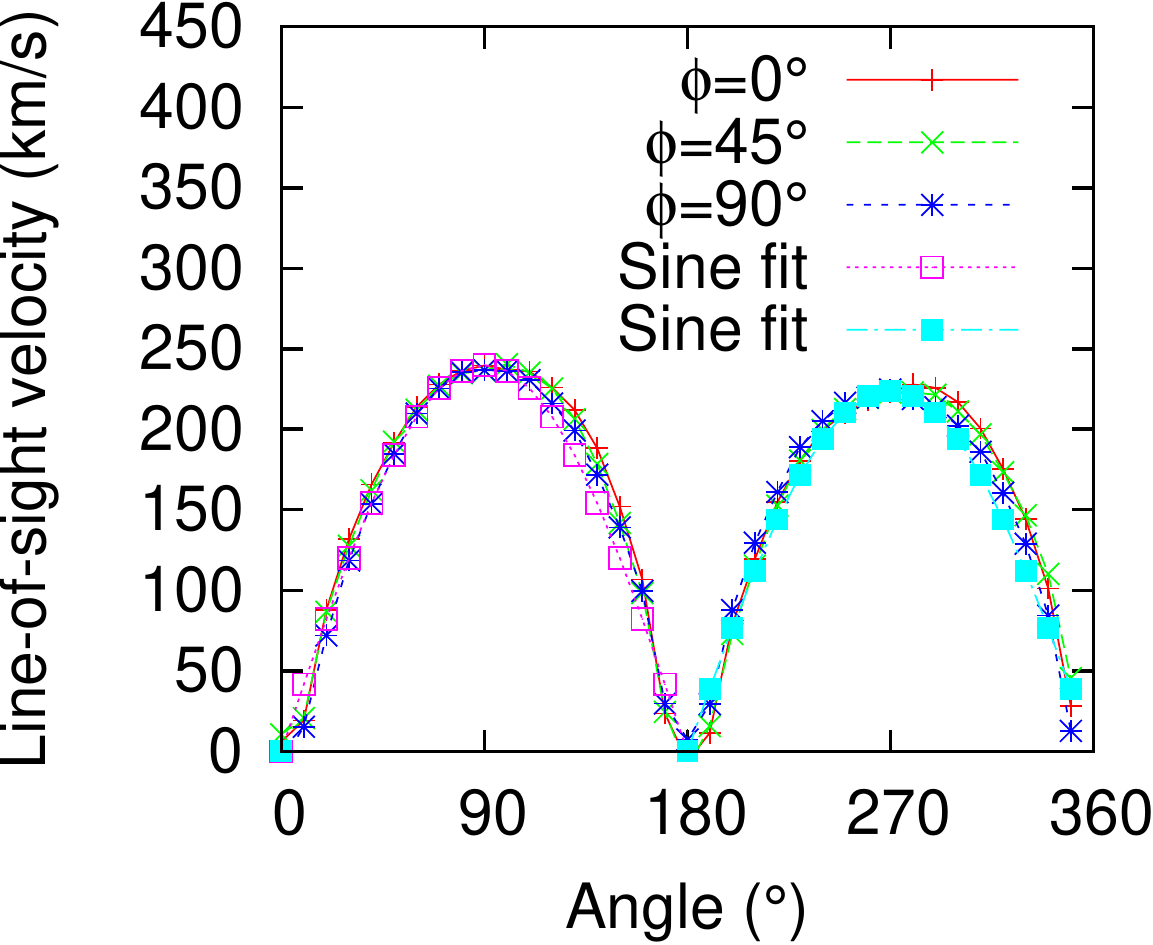}~\includegraphics[width=.48\columnwidth]{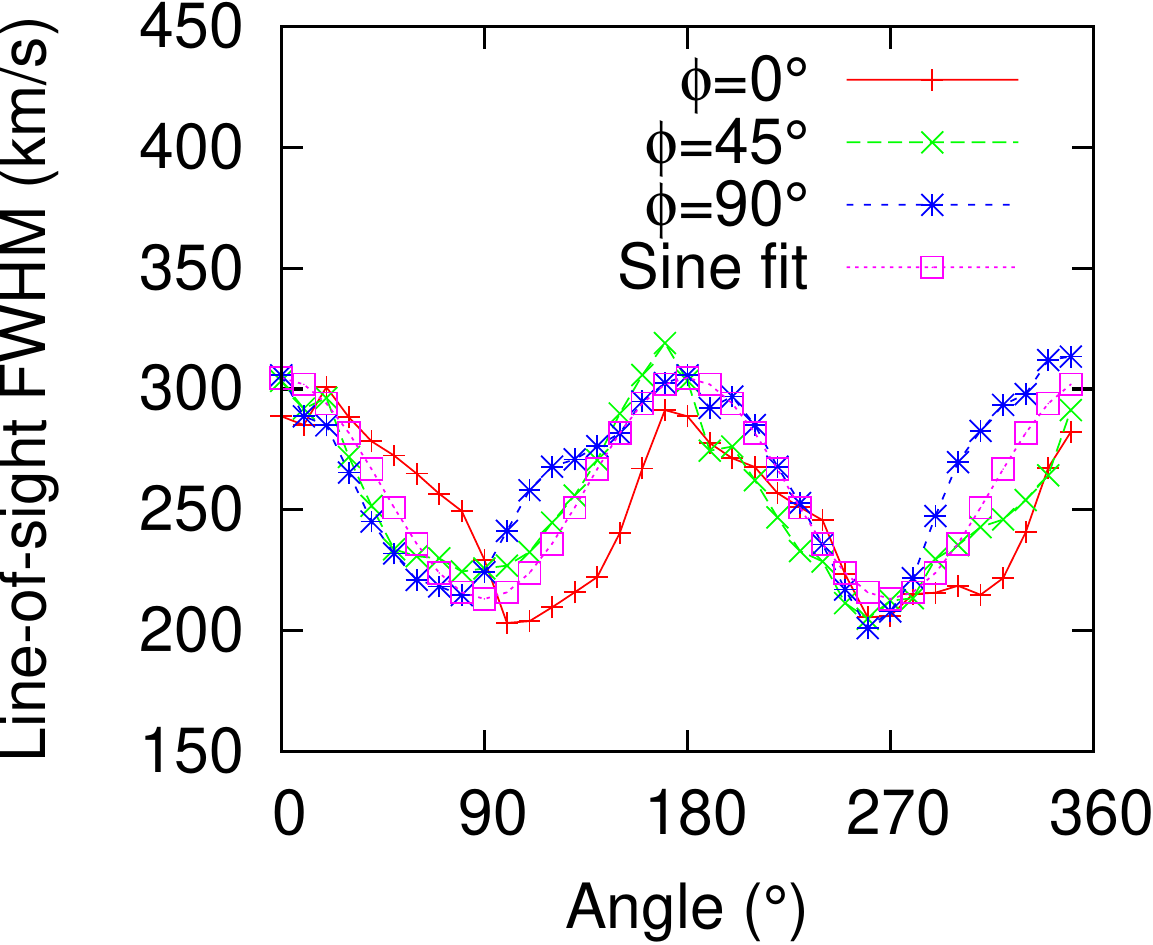}\\
\end{center}
\caption{\label{outflowv}
Left: Fitted outflow velocities at line-centre as a function of elevation angle for BHighRes.
Right: Line width as a function of elevation angle for three azimuthal angles
}
\end{figure}

It is clear from these plots that there is substantial broadening in the outflow. If we remove the thermal contribution to the broadening, replacing each Gaussian line profile with a top-hat function with width equal to our spectral resolution (i.e. the numerically resolution-limited equivalent to a Dirac delta function), then this broadening is still present (Fig.~\ref{thermalnonthermal}). This confirms that instabilites in outflows can produce cold gas with large non-thermal broadening.

The left frame of Fig.~\ref{outflowv} shows the line-centre outflow velocity as a function of viewing angle. The variation with elevation angle is closely modelled by an absolute sine function, which demonstrates that the outflow appears to be a single coherent flow. This is not surprising as the spatial resolution is essentially the entire disc, and so any small-scale deviations will be smoothed out. There is also no significant variation with azimuthal angle. This result is consistent with the line-widths below, and so it can be concluded that on a broad scale, this model is azimuthally symmetric. Note that this does not necessarily justify the accuracy of two-dimensional simulations: non-axisymmetric instabilities and flows are still dominant in the simulation (see Fig.~\ref{coldens0153} for example). Although these full-scale galaxy models evolve from axisymmetric initial conditions, the symmetry is broken through the stochastic feedback algorithm.

The right frame of Fig.\ref{outflowv} shows the full-width at half-maximum (FWHM) as a function of viewing angle. The line-width peaks at edge-on viewing angles, and is at a minimum for face-on viewing angles, approximately fitting a sine wave. This appears to show that most of the velocity dispersion is parallel to the plane of the disc. This may be seeded by the rotation of the disc, although the outflowing gas is no longer coherently rotating.

\subsubsection{Comparison with observations}\label{bgalobs}

The outflow velocities from the simulated spectra of the full galaxy simulations are significantly smaller than the average observed by \citet{2005ApJ...621..227M} of $330\pm100$ km/s. Our maximum (i.e. face-on) outflow velocity is $240$ km/s, which agrees with Martin's result though only at the limits of the allowed errors. At any other angle our outflow velocities are slower. The cause may be simply that the simulated star formation rate and galaxy mass are smaller than a typical ULIRG, or that we have neglected the energy impact from an AGN. However, there is also a significant scatter in the relationship between star formation rate and outflow velocity \citep{2005ApJ...621..227M}, and indeed there are observed ULIRGs with star formation rates of $\approx 400$ M$_\odot$/yr with outflow velocities of only $\approx100$ km/s.

\begin{figure}
\begin{center}
\includegraphics[width=1.\columnwidth]{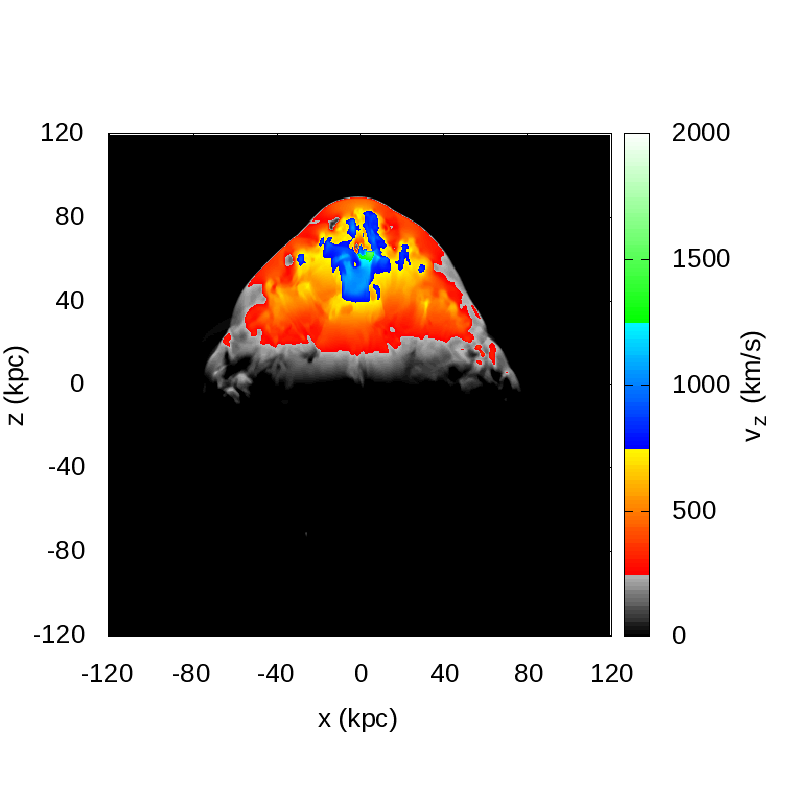}\\
\end{center}
\caption{\label{bigvely}
Vertical velocity (i.e. $v_z$) of BHighRes at $t=100$ Myr. The colour scheme is chosen to match Fig.~\ref{garishvelocity}.
}
\end{figure}

By comparison, the {\em hot} gas in this simulation reaches extremely high velocities, as high as $2000$ km/s, although most of the volume of gas still has velocities of less than $1000$ km/s (Fig.~\ref{bigvely}). While the clouds do not have significantly slower velocities than hot gas at the same height above $z=0$, the hot gas extends much further above the disc, and hence the maximum velocity of the hot gas greatly exceeds that of the cold clouds.

The line-widths range from $200$ km/s to $300$ km/s, which agrees with the lower range of \citet{2005ApJ...621..227M}'s observations of $320\pm120$ km/s (although there is significant scatter in the observations; \citet{2009ApJ...703.1394M} observe line-widths up to $800$ km/s). This shows that it is indeed possible for cold clouds in ULIRG outflows to have sufficient velocity dispersion to explain the large observed line-widths.

\begin{figure}
\begin{center}
\includegraphics[width=.5\columnwidth]{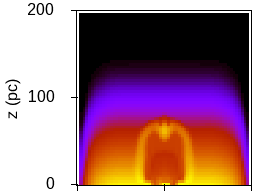}\includegraphics[width=.5\columnwidth]{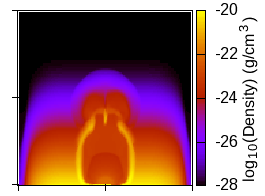}\\
\includegraphics[width=.5\columnwidth]{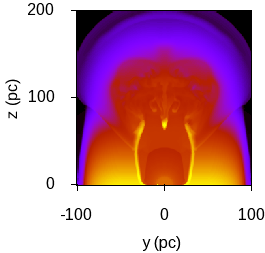}\includegraphics[width=.5\columnwidth]{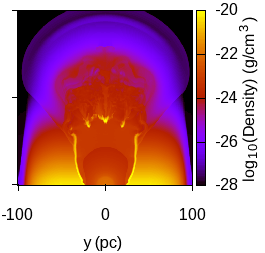}\\
\end{center}
\caption{\label{dens_res}
Density slices of galaxy centre models at various effective resolutions. Slices of models are taken at $130$ kyr. Top left: SVLowRes, effective resolution $64^3$. Top right: SLowRes, $128^3$. Bottom left: SMedRes, $256^3$. Bottom right: SHighRes, $512^3$.
}
\end{figure}

\begin{figure*}
\begin{center}
\includegraphics[width=.977\columnwidth]{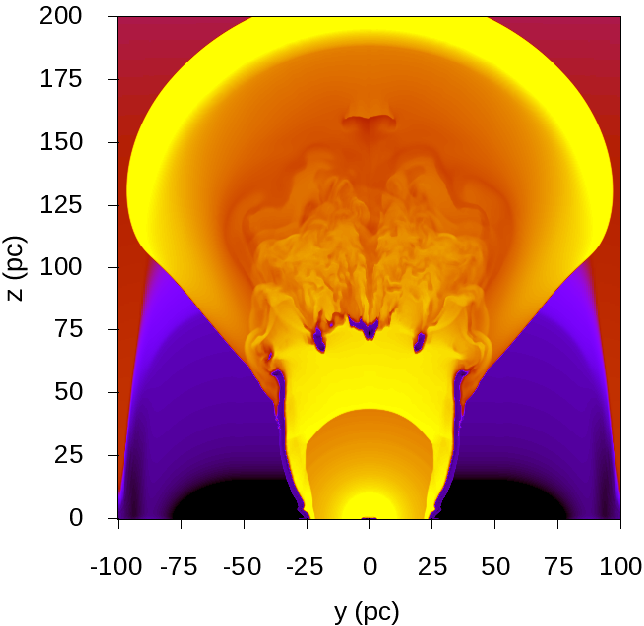}~
\includegraphics[width=1.023\columnwidth]{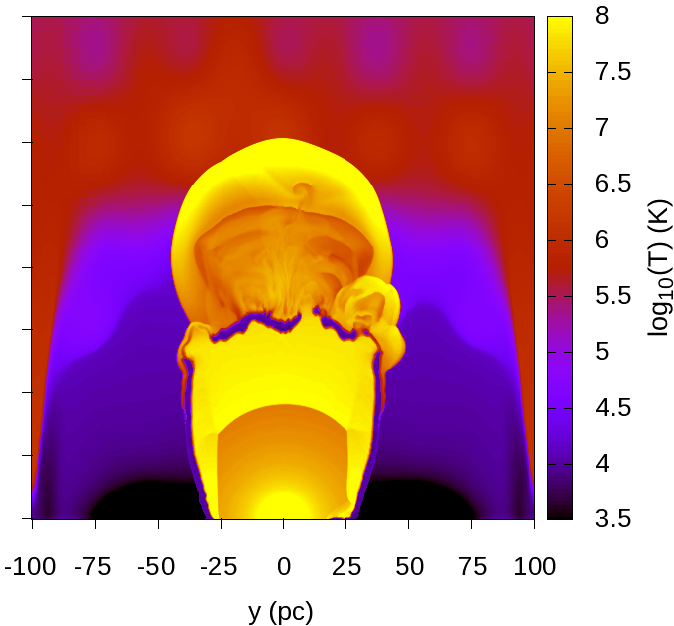}\\
\end{center}
\caption{\label{slice0154}
Edge-on slice of temperature for SHighRes (left) and SHighResLumpy (right) at $t=130$ kyr.
}
\end{figure*}

\begin{figure}
\begin{center}
\includegraphics[width=1.\columnwidth]{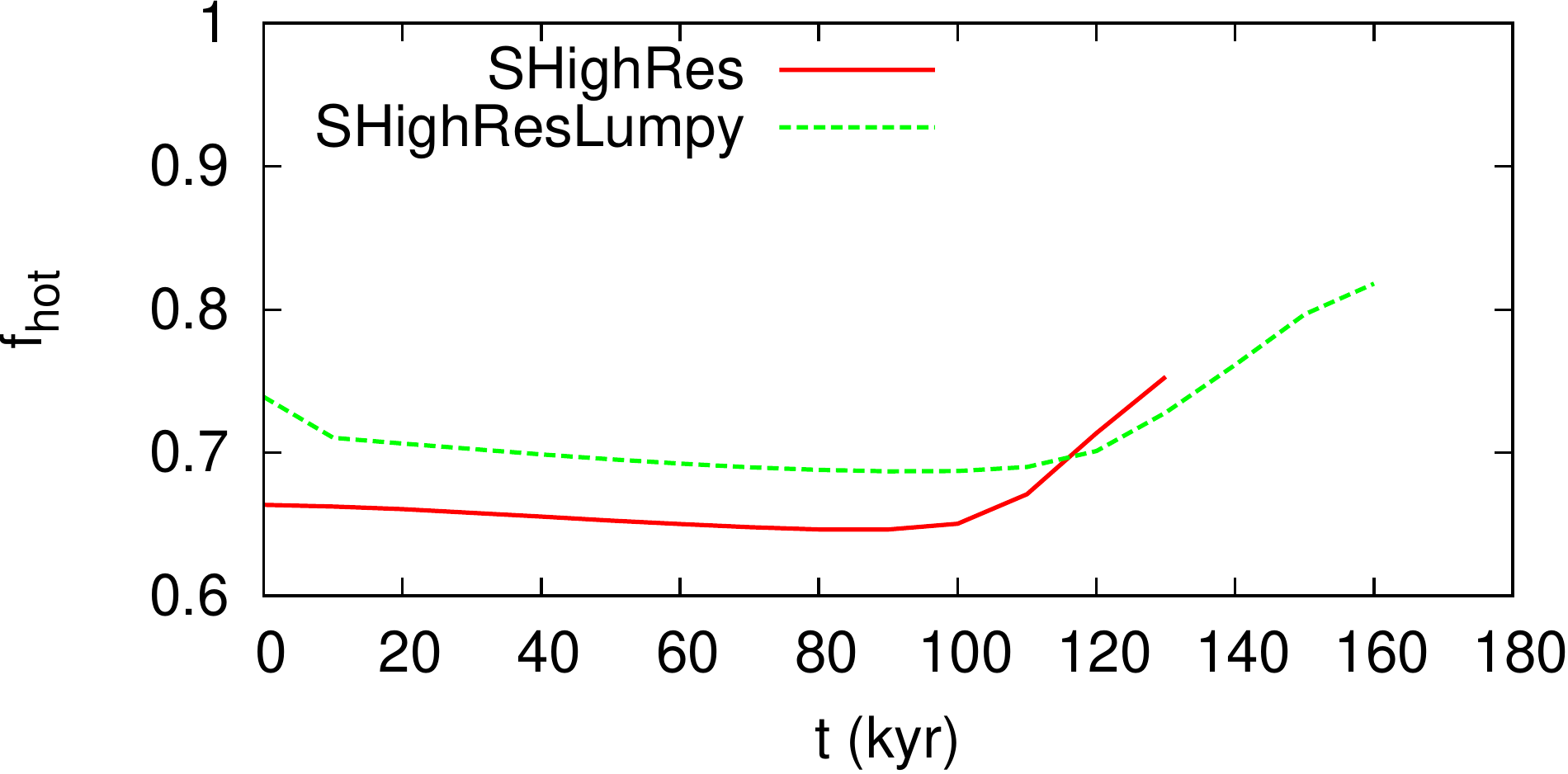}~ 
\end{center}
\caption{\label{hotlumpy}
{Evolution of the volume fraction $f_\mathrm{hot}$ of gas with $T>10^5$ K in SHighRes and SHighResLumpy. Initially $f_\mathrm{hot}$ is higher in  SHighResLumpy due to the initial density perturbations resulting in small regions of hot gas above the disc. In each model $f_\mathrm{hot}$ rapidly increases when the simulation reaches blow-out, at $t\approx100$ kyr for SHighRes, and $t\approx120$ kyr for SHighResLumpy.}
}
\end{figure}

\begin{figure}
\begin{center}
\includegraphics[width=1.\columnwidth]{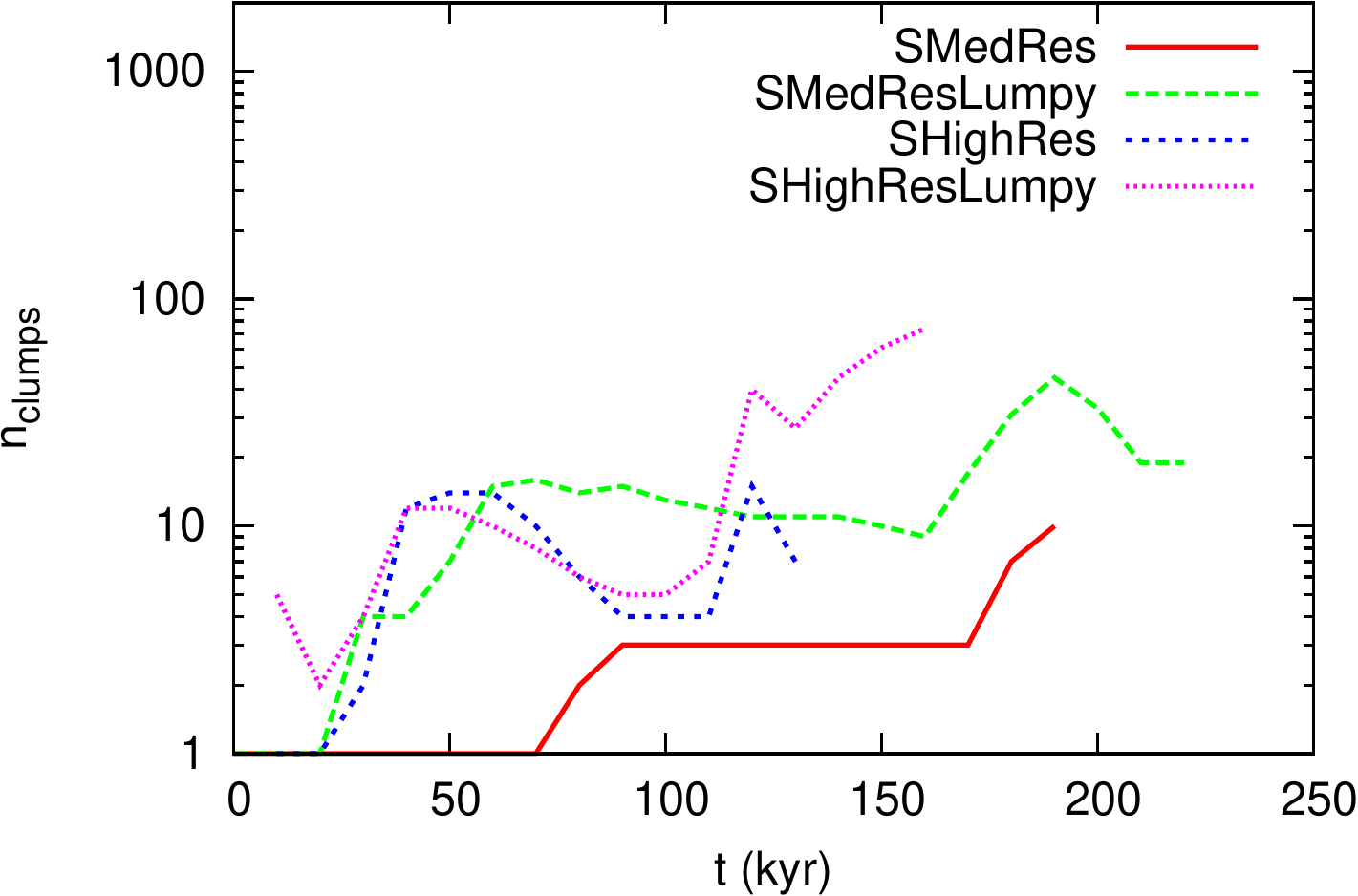}
\end{center}
\caption{\label{nclumpsturb}
Number of clouds in galaxy centre models.
}
\end{figure}

\subsection{Galaxy Centre Models}\label{galaxycentreresults}

\subsubsection{General evolution}

In these models, the centralized feedback inflates a bubble which rapidly rises through the disc, sweeping up dense matter until instabilities allow the hot gas to break through. At lower resolutions, the entire centre of the bubble ``blows out'', and no complex structure is formed. However, at higher resolutions the bubble wall fragments into a number of dense regions, around which the hot gas flows. These results are illustrated in Fig.~\ref{dens_res}. Here --- as in \citet{2009ApJ...698..693F} --- the origin of fragmentation does indeed appear to be the RT and RM instabilities, but the nature of the fragmentation differs greatly in our models.

Although at $t=130$ kyr, a 2D slice of SHighRes appears to show clouds forming from the bubble wall (left panel of Fig.~\ref{slice0154}), the cold gas is not a series of clouds but still a single contiguous structure. It appears that the hot gas has punched holes in the cold shock front, but that the walls surrounding these holes have remained largely intact. The structure of these walls are strongly symmetric, due to the symmetry of the feedback and initial conditions. The inclusion of asymmetric density perturbations in the initial conditions of SHighResLumpy breaks this pattern (right panel of Fig.~\ref{slice0154}), and produces a more irregular cloud structure. This ``blow-out'' also occurs a later time in SHighResLumpy (Fig.~\ref{hotlumpy}), perhaps as a result of increased cooling in the dense perturbations.

At this point in the simulation, hot gas can flow through the fragmenting bubble wall, filling much of the simulation domain with hot high-speed gas. This drives down the time-step and forces refinements across a greater volume of the simulation, dramatically increasing the wall clock time per output. As a result, there was insufficient wall-clock time to fully evolve SHighRes. However, we have allowed one high-resolution simulation -- SHighResLumpy -- to evolve further so that we can examine the evolution of the fragmenting bubble wall. Our fiducial line profiles are taken from SHighResLumpy at $t=160$ kyr.

\subsubsection{Formation and evolution of clouds}

We applied our cloud finding algorithm to these simulations (Fig.~\ref{nclumpsturb}). Most of the models did not have sufficient time to fragment into a large number of small clouds, but SHighResLumpy produced $>70$ clouds by the end of the simulated time. These clouds were still being strongly accelerated by the hot gas by the end of the simulation, and hence it is not appropriate to use the ballistic approximation to predict the trajectories. However, these clouds will still contribute to the non-thermal broadening of the NaI line, which we analyze in the following section.

\subsubsection{Estimates of neglected effects on clouds}\label{smallneglect}

Following the procedure outlined in section \ref{bigneglect}, we can estimate the effects of self-gravity and thermal conduction, as well as the importance of hot flows below our resolution in cloud destruction for the galaxy centre models. We select parameters that are typical of clouds and the wind in these simulation, setting $T_h\approx10^7$ K, $T_c\approx5\times10^4$ K, $n_c\approx600$ cm$^{-3}$, $\Lambda\approx10^{-22}$ erg cm$^3$ s$^{-1}$, $l_c\approx0.3$ pc, $v_c\approx 1300$ km/s, $v_h\approx2000$ km/s, $\rho_c\approx10^{-21}$ g cm$^{-3}$, and $\rho_h\approx10^{-24}$ g cm$^{-3}$.

\begin{figure}
\begin{center}
\includegraphics[width=.98\columnwidth]{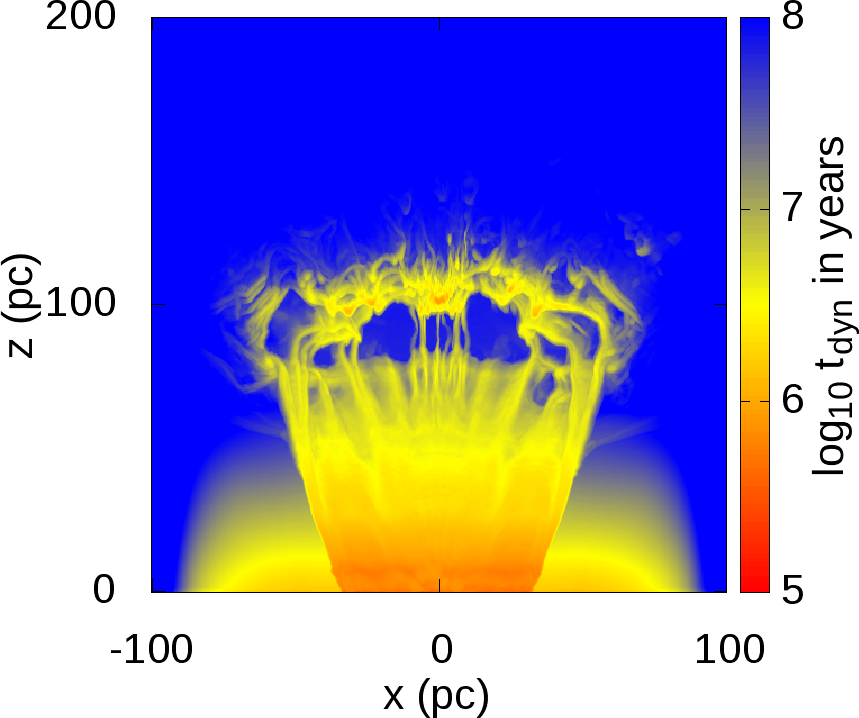}\\
\end{center}
\caption{\label{tdyn0185}
{The minimum dynamical times along the line-of-sight for an edge-on view of SHighResLumpy at $160$ kyr.} 
}
\end{figure}

\begin{figure*}
\begin{center}
\includegraphics[width=.962\columnwidth]{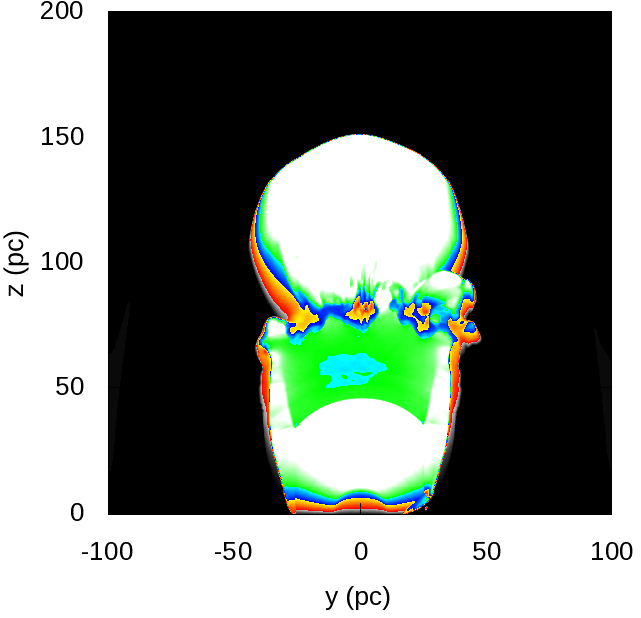}\includegraphics[width=1.038\columnwidth]{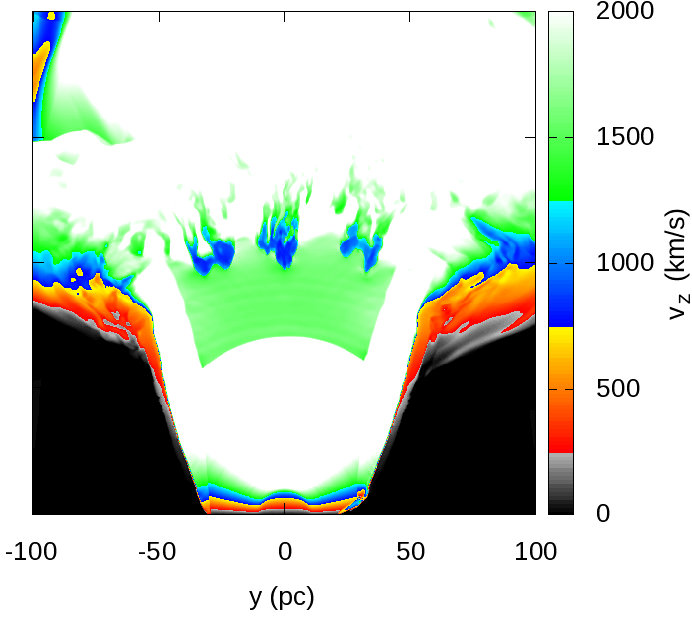}\\
\end{center}
\caption{\label{garishvelocity}
Vertical velocity (i.e. $v_z$) of SHighResLumpy at $t=130$ kyr (left), and $t=160$ kyr (right). The colour scheme is chosen to divide the gas into four categories: $v\approx0$ km/s, $v\approx500$ km/s, $v\approx1000$ km/s, and $v\gg1000$ km/s.
}
\end{figure*}

The minimum dynamical times along the line-of-sight for an edge-on view SHighResLumpy at $160$ kyr is plotted in Fig.~\ref{tdyn0185}. The smallest dynamical times above the disc are $t_\mathrm{dyn}\approx1$ Myr, and so self-gravity is negligible in the evolution of these clouds.

The classical thermal conduction rate of $\nabla\cdot q\approx1.8\times10^{-18}$  erg s$^{-1}$ cm$^{-3}$, and a radiative cooling rate of $\Lambda n_c^2\approx3.6\times10^{-17}$ erg s$^{-1}$ cm$^{-3}$, and so conduction is not dominant. The mean-free path for hot electrons is $0.5$ pc, on the order of the size of the clouds, suggesting that the correct thermal conduction rate may be a little less than this classical limit.

The cloud crushing time for the central clouds is $\approx140$ kyr, similar to the simulation time. Hence we should indeed expect these clouds to continue to be disrupted as the system evolves. In paper II we will make use of a sub-grid model that incorporates these effects.

\subsubsection{Simulated spectra}\label{specsmall}

We applied our raytracing code to produce simulated spectra of the models SHighResLumpy. In contrast to the full-galaxy run, in this model the initial disc is largely intact outside of the central feedback region. Furthermore, even after blow-out, the lower portions of the bubble wall remain intact. These regions are both dense and cold, and contain NaI. An unfiltered face-on spectrum with an effective spatial resolution broad enough to cover the entire simulation shows three components: a sharp line at $v\approx0$ km/s, a broad line at $v\approx500$ km/s and a weaker line at $v\approx1000$ km/s. Fig.~\ref{garishvelocity} shows the z-velocities of slices through SHighResLumpy, with a colour scheme chosen to emphasize these three broad regions of velocity. The $v\approx0$ km/s line clearly corresponds to the intact disc, while the $v\approx500$ km/s line corresponds largely to the outer portions of the burst-open bubble wall. Only the $v\approx1000$ km/s line corresponds to fragmenting material within the outflow. As this work is primarily focused on the outflowing gas, the disc component can be neglected, and we do this by cutting out all gas below $z=70$ pc. The lower components of the wall have significant velocities ($\approx500$ km/s), but this dense gas is not entrained within the hot flow, and is not being accelerated by it. As shown in Fig.~\ref{garishvelocity}, the clouds are accelerated from $\approx500$ km/s to $\approx1000$ km/s from $t=130$ kyr to $t=160$ kyr, while the lower walls remain at a roughly constant speed. Hence the gas of interest only consists of the clouds near the centre of the outflow. To cut out this outer wall, the beam-width is narrowed to $60$ pc.

\begin{figure}
\begin{center}
\includegraphics[width=.48\columnwidth]{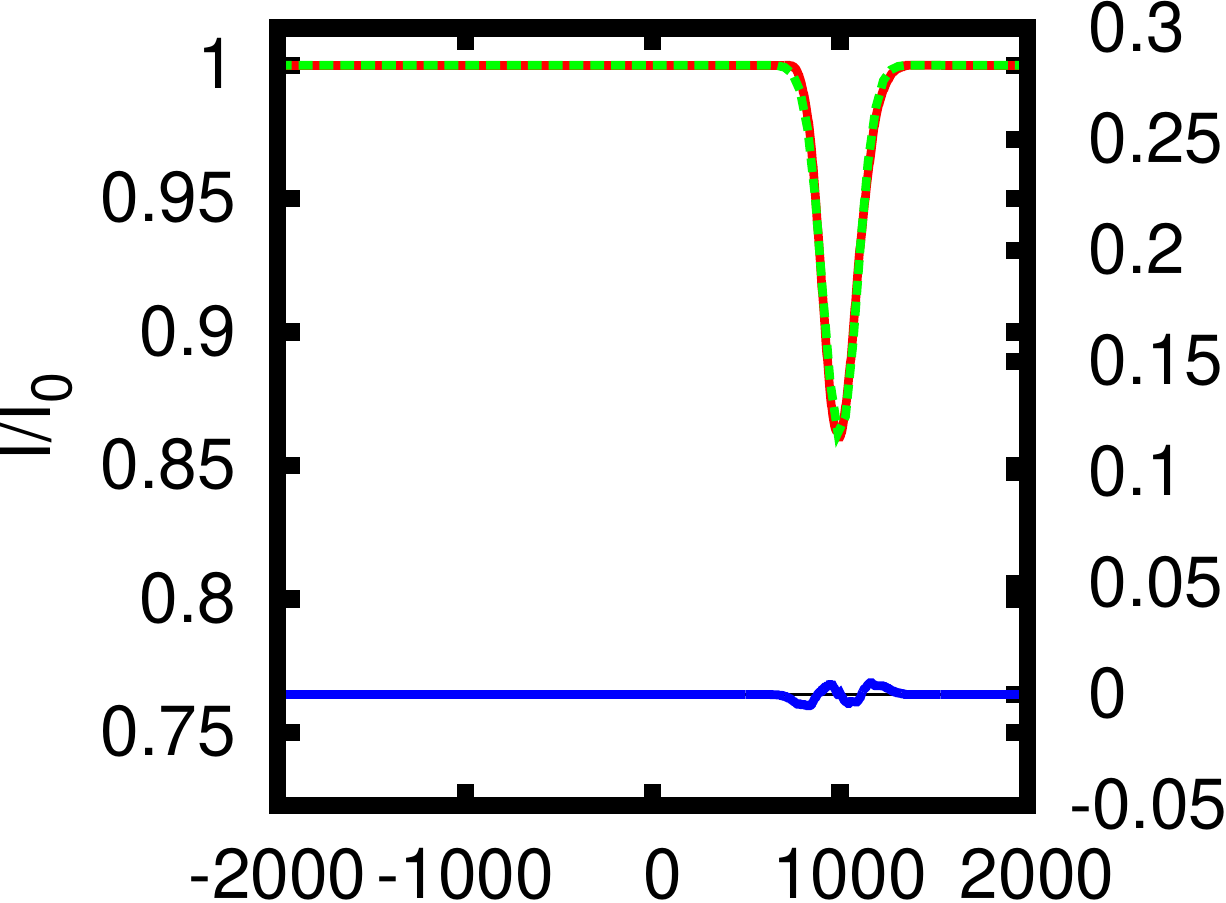}~\includegraphics[width=.48\columnwidth]{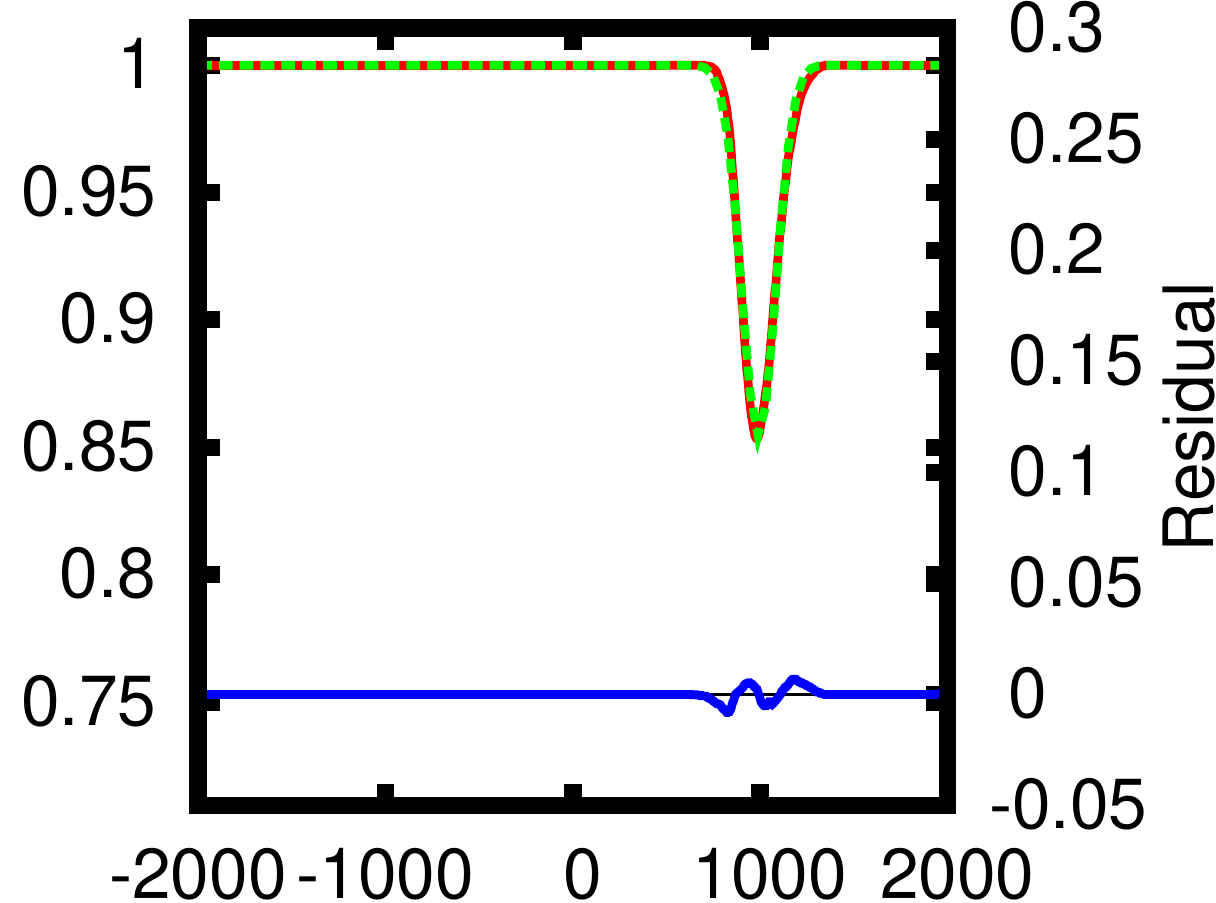}\\
\includegraphics[width=.48\columnwidth]{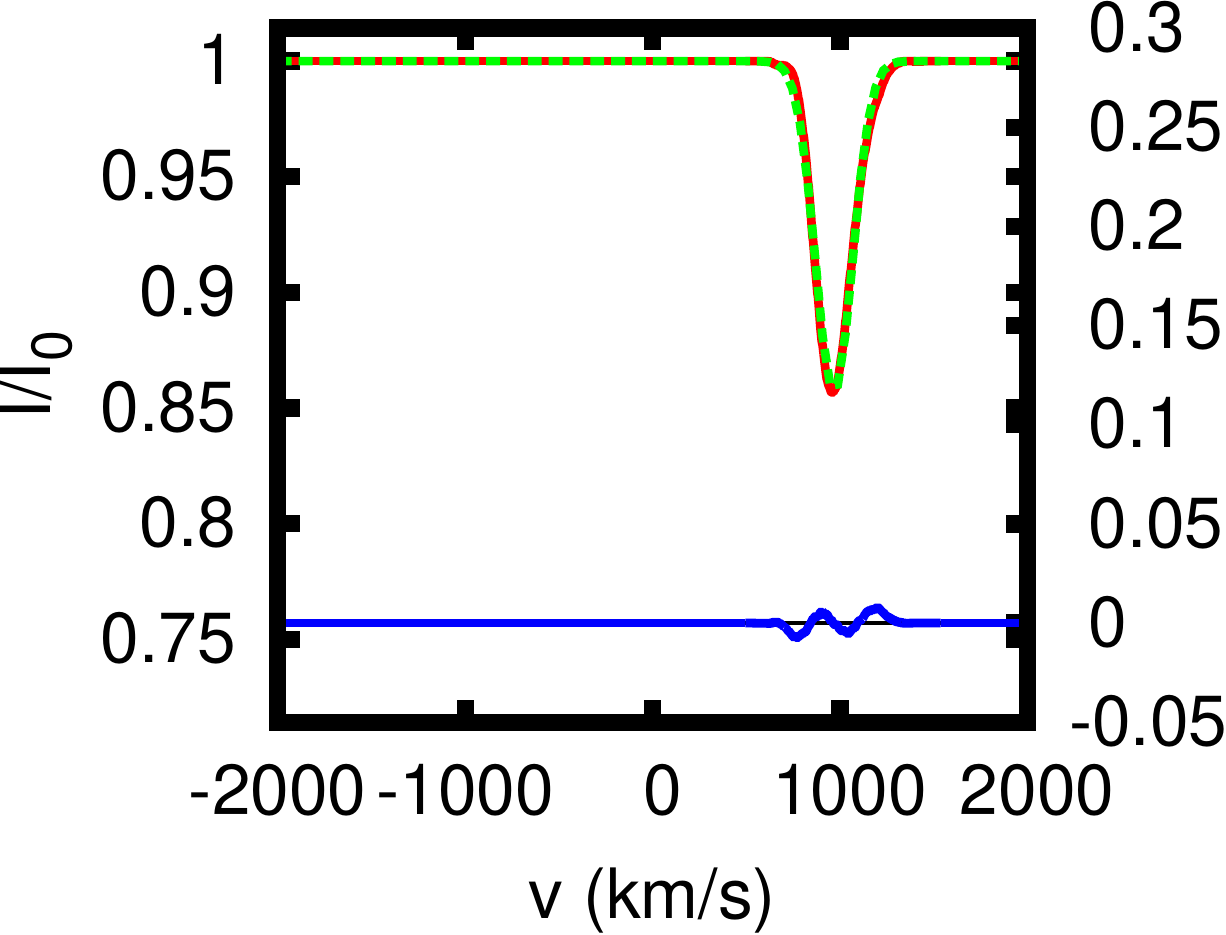}~\includegraphics[width=.48\columnwidth]{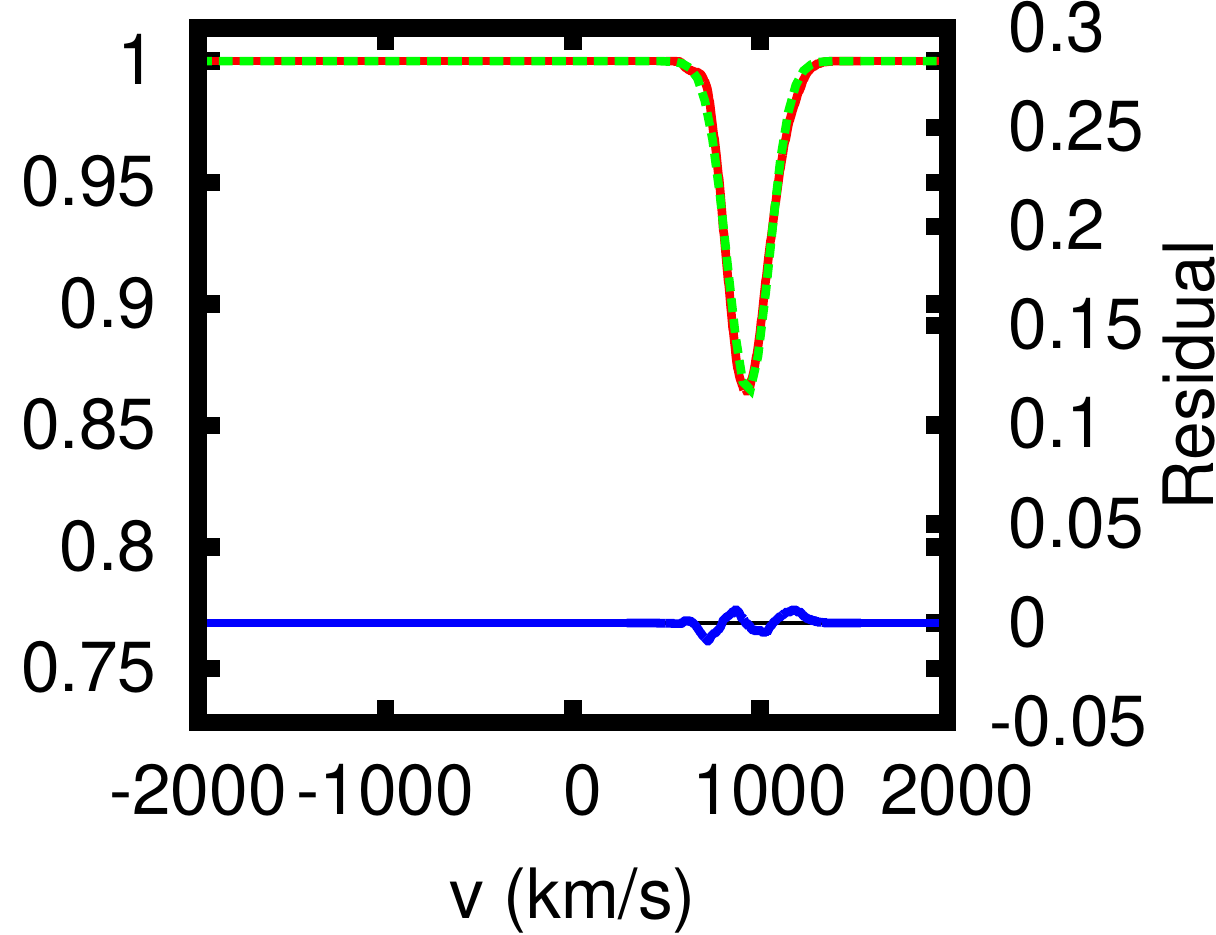}

\end{center}
\caption{\label{centrallines}
Line profiles for angles of $0^\circ$ (top-left), $6^\circ$ (top-right), $14^\circ$ (bottom-left), and $20^\circ$ (bottom-right) from face-on, for the simulation SHighResLumpy. The red line is the line profile, while the green line is a Gaussian fit, both described by the y-axis labels on the left of each plot. The blue line is the residual, described by the y-axis labels on the right of each plot.
}
\end{figure}

\begin{figure}
\begin{center}
\includegraphics[width=.48\columnwidth]{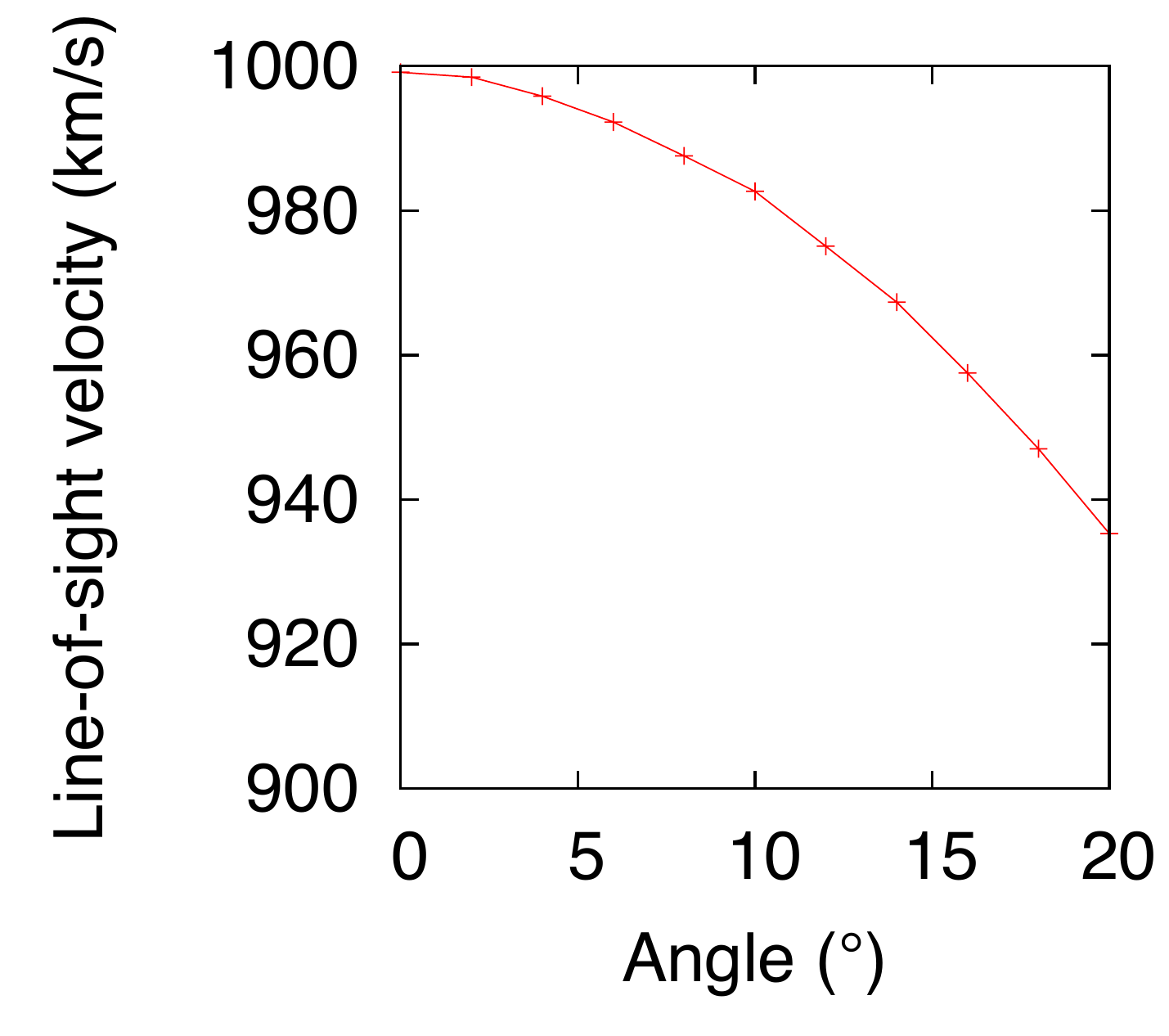}~\includegraphics[width=.48\columnwidth]{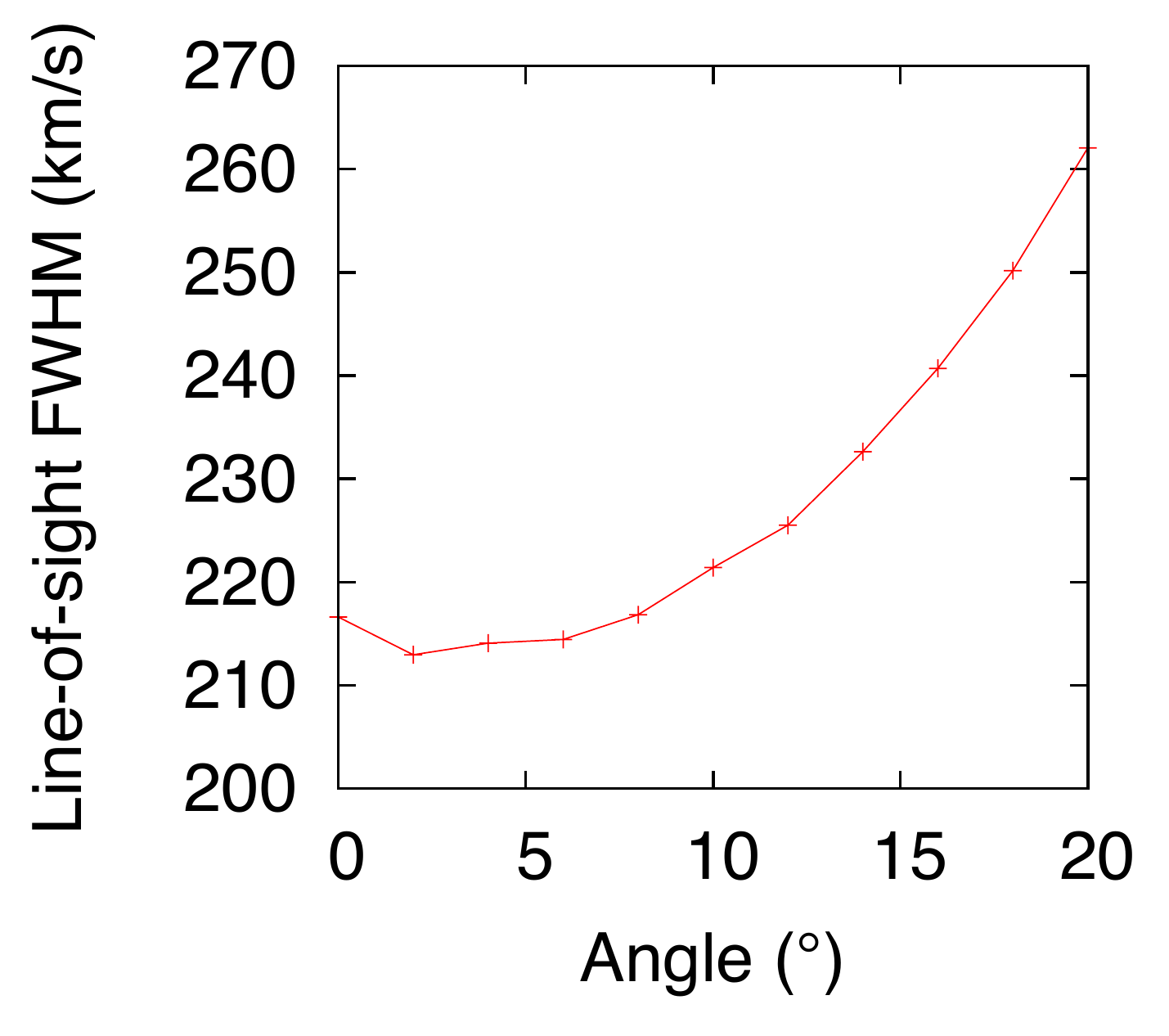}\\
\end{center}
\caption{\label{angleoutflowcentral}
Line-of-sight velocity at line centre (left), and line-width (right) as a function of viewing angle for SHighResLumpy, with the disc removed and a spatial resolution that resolves the outflow while missing the lower bubble wall.
}
\end{figure}

Fig.~\ref{centrallines} shows the line profiles from rays through SHighResLumpy at $t=160$ kyr at angles of $0-20^\circ$ from face-on. It is not useful to fire rays at angles greater than this, as the still intact lower bubble wall would dominate the spectrum. The velocity dispersion of the clouds produces a curve that is well approximated by a Gaussian. The lines show an outflow velocity of up to $1000$ km/s and a FHWM of up to $220$ km/s. There is no strong difference between the line profiles as the viewing angle is changed by this small amount. However, as shown in Fig.~\ref{angleoutflowcentral}, the velocity at line-centre decreases and the velocity dispersion increases as the viewing angle moves away from a direct face-on view. This is unsurprising, as the main component of the net velocity is in the vertical direction, while the main component of the velocity dispersion is in the horizontal direction.

To test the effects of the asymmetric perturbations, we also determined spectra for SHighRes and SHighResLumpy at $t=130$ kyr. At this earlier time, the outflow velocities are lower ($800$ km/s and $850$ km/s), as the clouds have had less time to accelerate, but are similar between the two simulations. However, the line-depth ($1-I/I_0$) at lines centre of $0.4$ is much deeper for SHighResLumpy at $t=130$ kyr than for SHighRes at $t=130$ kyr or SHighResLumpy at $t=160$ kyr, which both have line-depths of $0.14$. As stated above, blow-out occurs later in SHighResLumpy, and so the clouds are less disrupted and more optically thick at $t=130$ kyr.

\subsubsection{Comparison with observations}

Interestingly, despite the large difference in initial conditions and system evolution, the FWHM of the spectral lines from central-disc model are similar to the full galaxy case, both reaching a minimum of around $220$ km/s for a face-on view. These agree with the lower limit of \citet{2005ApJ...621..227M}'s value of $330\pm100$ km/s, and so we have confirmed that the break-up of the bubble wall does indeed produce sufficient velocity dispersion to explain the observed broadening. However, here the outflow velocity ($\approx1000$ km/s) greatly exceeds the observed velocities of $330\pm100$ km/s. This may simply be a product of the feedback conditions --- we may have overestimated the energy input rate, or underestimated the size of the feedback region, producing unrealistically intense feedback.

\subsubsection{Mass-loading}\label{massloading}

We also examined the effects of varying the mass-loading rate. As mentioned above, this is parameterised by $M_\mathrm{SN}$, the mass of gas ejected in each supernova. Modifying this value is equivalent to a combination of (a) varying the initial mass function of massive star formation, (b) including a low mass-loading AGN, and/or (c) varying the amount of gas that completely escapes a supernova. We have produced simulations with $M_\mathrm{SN}=0 \mathrm{M_\odot}$, $M_\mathrm{SN}=4 \mathrm{M_\odot}$, and $M_\mathrm{SN}=16 \mathrm{M_\odot}$ in addition to the fiducial $M_\mathrm{SN}=6$ M$_\odot$. These were performed at a low resolution, and so it is not possible to perform a detailed analysis of the variation in cloud formation and properties with mass-loading. However, we can investigate the impact on bubble evolution and blow-out. This is a scenario that has been well-studied in the past \citep[e.g.][]{1989ApJ...337..141M,1989ApJ...345..372N,1991ApJ...375..239F,1992ApJ...388..103K,2009ApJ...701..330S,2012MNRAS.425.2343Z}.

The primary result is that bubble inflation and blow-out occur much more rapidly and violently at smaller mass-loading rates. During blow-out, the volume  of hot gas increases dramatically, and so we can plot the volume fraction of hot gas in the models over time to compare the times at which blow-out occurs. With ``hot gas'' defined as gas above $10^5$ K, this is plotted in Fig.~\ref{hotgasres}, and it is clear that blow-out is weakened and delayed with increasing mass-loading (the initial gentle downwards slope is caused by cooling). The addition of mass raises the density within the bubble, which decreases the cooling time, reducing the effectiveness of heating. A greater mass also requires more kinetic energy to reach a high velocity. This also reduces the density contrast across the bubble interface, weakening the Rayleigh-Taylor instability.

\subsubsection{Resolution Dependence}\label{resdepends}

\begin{figure}
\begin{center}
\includegraphics[width=1.\columnwidth]{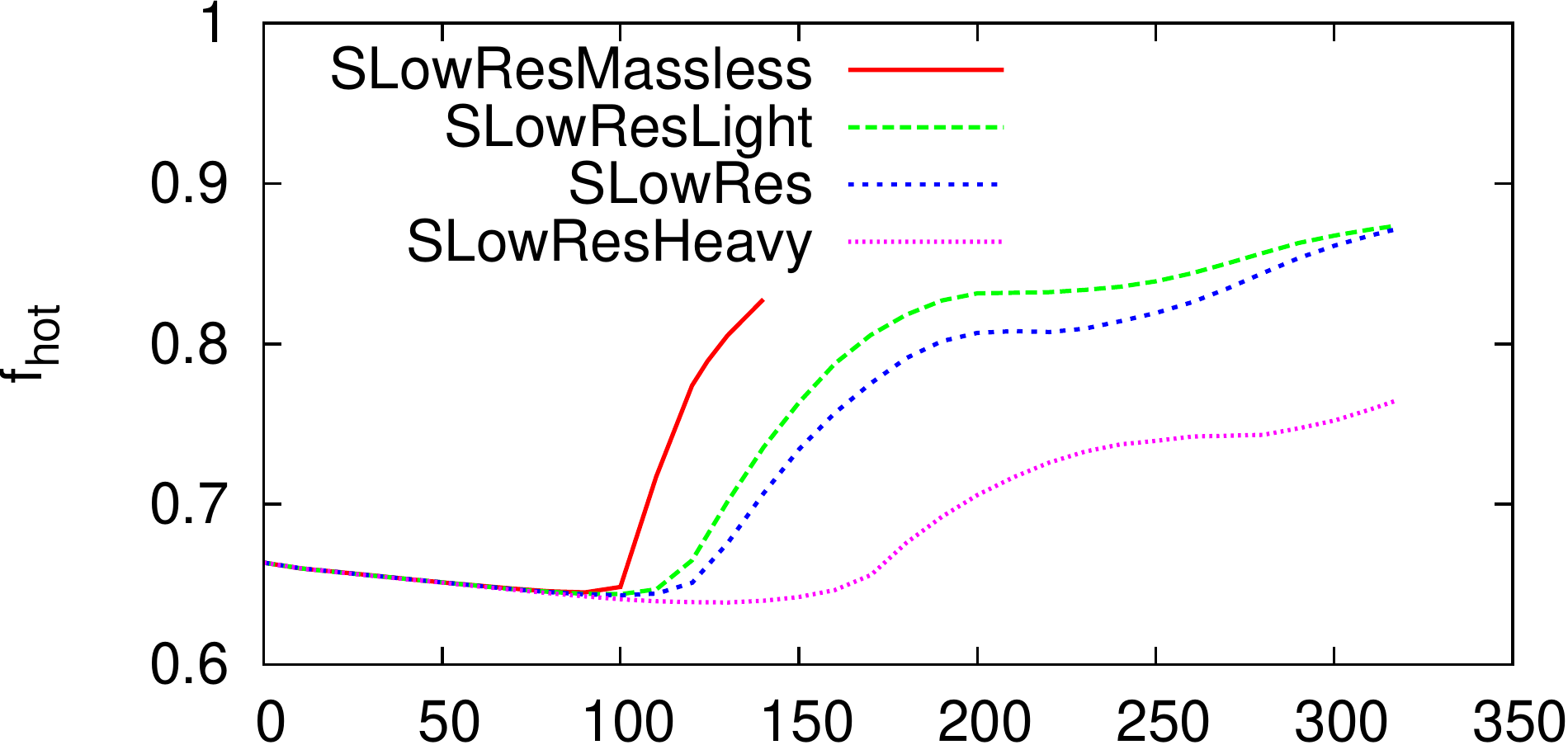}\\
\includegraphics[width=1.\columnwidth]{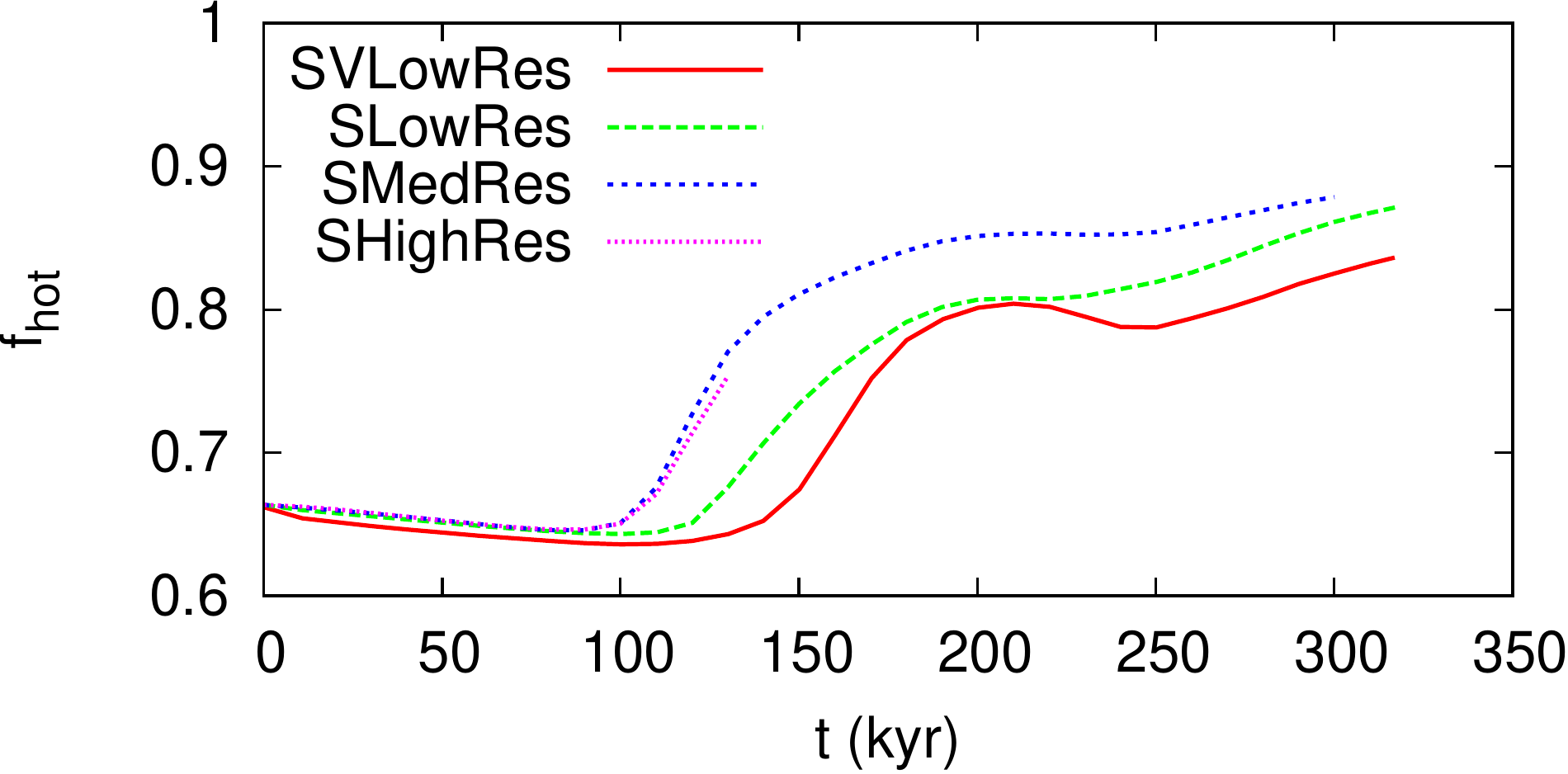}\\
\end{center}
\caption{\label{hotgasres}
Top: Evolution of hot gas volume fraction as mass-loading is varied. Bottom: Evolution of hot gas volume fraction as resolution is varied. The drop in hot gas fractions at the end of some simulations is due to hot gas escaping the simulated region.
}
\end{figure}

We examined models with four different maximum levels of refinement, with effective resolutions ranging from $64^3$ to $512^3$. The effect on blow-out times is clear in the hot gas fraction plots, Fig.~\ref{hotgasres}. As in section~\ref{massloading}, the hot gas fraction initially decreases slowly as gas radiatively cools throughout the disc, and then dramatically rises when ``blow-out'' occurs and hot gas begins to stream out of the bubble. A resolution dependence is clear: blow-out occurs at earlier times as resolution is increased. This is likely because the growth-rate of the RT instability increases with decreasing wavelength --- for an inviscid medium, the growth rate of an instability with amplitude $\eta$ in a gravitational field of acceleration $g$ is given by

\begin{equation}
\frac{d\eta}{dt} = (Agk)^{1/2} \eta,
\end{equation}
where $A$ is the Atwood number, and $k=2\pi\lambda$ is the wave-number of the instability \citep{1961hhs..book.....C}. As resolution is increased, the stronger instabilities at shorter wave-lengths are no longer suppressed by the discretization of the grid, and the bubble-wall can fragment and allow blow-out earlier. Our medium and high resolution runs appear to follow the same locus, but this does not mean the simulation has converged --- while blow-out occurs at a similar time, the level of fragmentation is clearly different (Fig.~\ref{dens_res}).
\begin{figure}
\begin{center}
\includegraphics[width=.5\columnwidth]{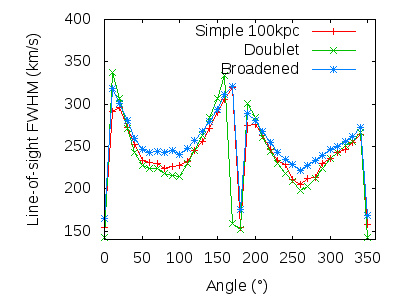}~\includegraphics[width=.5\columnwidth]{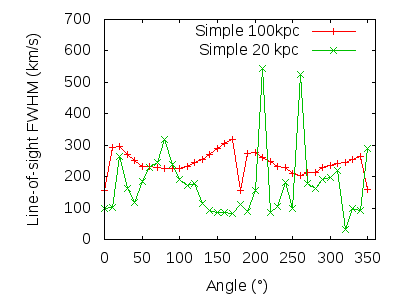}\\
\includegraphics[width=.5\columnwidth]{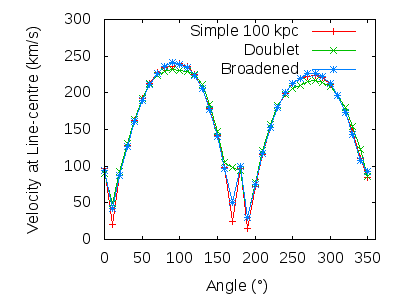}~\includegraphics[width=.5\columnwidth]{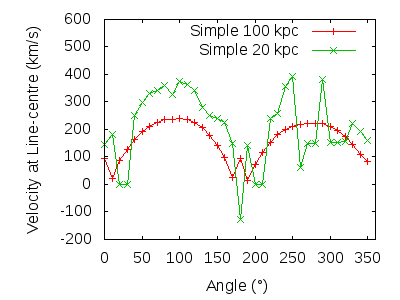}\\
\end{center}
\caption{\label{varylineprams} 	
The effects of extensions to the line profile models (left), and the effects of narrowing the ``beam-width'' (right).
}
\end{figure}

\subsection{Line profiles: Extended models}

We examined the effects of a more ``realistic'' line profile on BHighRes, including broadening from limited spectral resolution, and line confusion from inclusion of both lines in the Na D doublet. We also explored the results of using a much tighter spatial resolution (i.e. beam width), using $10$ kpc instead of our fiducial $100$ kpc. The results of these examinations are plotted in Fig.~\ref{varylineprams}.

The effective broadening due to spectral resolution is produced by convolving the intensity with a Gaussian curve, with a FWHM set to $65$ km/s. The second line in the doublet is modelled as a copy of the first line, transposed by $\Delta\lambda=5.97$ \AA. It is clear (left panels of Fig.~\ref{varylineprams}) that these additions do not significantly alter the line profile properties.

However, tightening the effective spatial resolution of the simulated observer {\em does} significantly change the line properties (right panels of Fig.~\ref{varylineprams}). A tight beam does not capture the details of the structure of the outflow --- the beam will likely intersect with only a few clouds, or none at all. This produces irregular results for the FWHM and outflow velocities that are not indicative of the overall structure of the outflow. While the very broad effective spatial resolution used in our fiducial models smooths out the large-scale structure, a tight beam misses this structure entirely. This confirms that, given the resolution limits of our hydrodynamic simulations, the ``broad beams'' used in our fiducial raytracing calculations are the most appropriate choice.

\section{Conclusions}\label{coldflowsconc}

We performed three-dimensional simulations to explain the source of cold high velocity-dispersion gas in ULIRG outflows \citep[as observed by][in particular]{2005ApJ...621..227M}. Our initial conditions were set up to produce a scenario where clouds are produced by the fragmentation of the wall of a galactic super-bubble, induced by the Rayleigh-Taylor instability \citep[as in][]{2009ApJ...698..693F}. This was done in two scenarios, one focusing on the central $200$ pc of the galaxy, and another where the entire galaxy is included in the simulation domain. We produced spectra of the simulations with a raytracing algorithm to facilitate comparison with observation.

Our models succeeded at producing cold outflowing gas with large velocity dispersions, but only at the higher resolutions. The velocity dispersions were $220-260$ km/s in our galaxy centre models, and $200-300$ km/s in our full galaxy models. These results are similar, and both agree with the lower end of the observations ($330\pm100$ km/s). Our two scales of simulation produce this cold outflowing gas through different means. In the highest resolution full galaxy models the cold disc gas is fragmented by the large number of hot bubbles produced in the disc. This gas is then pushed into the outflow by the intense pressure of the feedback beneath it, and is supplemented by cooling within the hot wind. In the lower resolution full-galaxy models, cold gas is only produced by cooling within the hot wind, or at very low resolutions, is not produced at all.

In the galaxy centre models, the cold gas is instead produced by the Rayleigh-Taylor fragmentation of the wall of a feedback-inflated bubble. Precise determination of the most important process will require larger-scale and higher resolution simulations, perhaps with improvements to the physical model such as the inclusion of self-gravity and a more self-consistent feedback algorithm. In paper II, we will perform simulations with a sub-grid turbulence model to better model feedback and alleviate resolution issues. Nevertheless, in this paper we have demonstrated that cold gas with high velocity-dispersions can indeed be present in simulations of ULIRG outflows.

\section*{Acknowledgments}

DJW is supported by the Natural Sciences and Engineering Research Council of Canada. RJT is supported by a Discovery Grant from NSERC, the Canada Foundation for Innovation, the Nova Scotia Research and Innovation Trust and the Canada Research Chairs Program. ES is supported by NSF grant AST11-03608. MB acknowledges support by the research group FOR 1254 funded
by the Deutsche Forschungsgemeinschaft. Simulations were run on the CFI-NSRIT funded {\em St Mary's Computational Astrophysics Laboratory}, on the {\em Arizona State
University Advanced Computing Center}, and on the Calcul-Qu\'{e}bec/Compute-Canada supercomputer {\em Colosse} at Universit\'{e} Laval. Simulations were performed using the FLASH code, a product of the DOE ASC/Alliances funded Center for Astrophysical Thermonuclear Flashes at the University of Chicago. We also thank our anonymous referee for suggestions that improved the content and presentation of this paper.

\bibliographystyle{mnras}
\bibliography{coldflows_noturb1_CLEANED}

\bsp

\label{lastpage}

\end{document}